\documentclass{svjour3}                     
\smartqed  

\usepackage{graphicx}
\usepackage{epstopdf} 
\usepackage{amsmath,amssymb}
\usepackage{natbib}
\usepackage{enumerate}
\usepackage{threeparttable}
\usepackage{booktabs}
\usepackage[dvipsnames]{xcolor}
\usepackage{hyperref}

\hypersetup{
    colorlinks=true,
    linkcolor=blue,
    citecolor=blue,
    filecolor=red,
    urlcolor=blue,
    pdftitle={Distance Estimates to Five Bok Globules},
    pdfauthor={Rajat Subhra Paul, Himadri Sekhar Das}
}

\journalname{Astrophysics and Space Science}

\begin{document}

\title{Distance Estimates to Five Bok Globules Using Gaia DR3 Parallaxes and Near-Infrared Photometry: Validation with 3D Dust Maps}

\titlerunning{Distance Estimates to Five Bok Globules}

\author{Rajat Subhra Paul \and Himadri Sekhar Das}

\institute{R. S. Paul \at
              Department of Physics, Karimganj College, Sribhumi 788711, India \\
              \email{apurajatsubhra@gmail.com} 
           \and
           H. S. Das \at
              Department of Physics, Assam University, Silchar 788011, India \\
              \email{himadri.sekhar.das@aus.ac.in} 
}

\date{Accepted for publication in Astrophysics and Space Science
}

\maketitle

\begin{abstract}
Accurate distances to small, opaque Bok globules are difficult to obtain due to their compact sizes and lack of embedded standard candles. We estimate distances to five globules---CB4, CB24, CB56, CB60, and CB188---using a near-infrared (NIR) extinction technique applied to 2MASS photometry, combined with stellar distances from \textit{Gaia} DR3 parallaxes. Extinction--distance profiles reveal distinct rises marking each cloud, yielding median distances of 766~pc (CB4), 354~pc (CB24), 483~pc (CB56), 1143~pc (CB60), and 926~pc (CB188). Weighted-mean values agree within $8.0\%$, supporting the internal consistency of the method and interquartile‑range-based uncertainties span 3--9\% across the sample. Independent validation with Bayestar19 3D dust maps supports the CB4 and CB188 results. These refined distances provide an improved basis for future determinations of cloud masses, densities, and star-formation efficiencies, including the first robust measurement for CB56 and a revised upward distance for CB188.

\end{abstract}

\section{Introduction}
\label{sec:intro}

Bok globules are small, nearly isolated molecular clouds with typical diameters of $0.1\text{--}2~\text{pc}$ and masses between $2$ and $100~M_{\odot}$ \citep{Bok1977}. These objects serve as ideal laboratories for investigating early-stage star formation, protostellar collapse, and magnetic field interactions \citep{Clemens1991}, while remaining relatively uncontaminated by nearby stellar clusters. Evidence of ongoing low-mass star formation includes bipolar molecular outflows \citep{Yun1994a}, sub-millimeter continuum emission, and infrared colors consistent with Class 0 or Class I protostars \citep{Yun1994b, Launhardt1997}.

Accurate distance measurements are fundamental to deriving key physical parameters such as mass, size, density, and luminosities of embedded young stellar objects (YSOs) \citep{Clemens1991, Yun1990}. Without precise distances, it is impossible to distinguish genuine cloud properties from apparent differences arising from projection effects or distance uncertainties. Distance determination evolved from early star count methods \citep{Wolf1923, Bok1941} and bracketing techniques \citep{Hobbs1986} to modern approaches utilizing broadband photometry \citep{Peterson1998, Hearty2000}, extinction mapping \citep{Lombardi2008}, and astrometric missions. The \textit{Gaia} mission provided unprecedented parallax precision \citep{Gaia2016, Gaia2022}, enabling revolutionary improvements in cloud distance determinations.

The near-infrared (NIR) photometric technique, successfully applied by \citet{Maheswar2010} using 2MASS ($J\text{--}H$) and ($H\text{--}K_{S}$) color indices of main-sequence stars (A0–K7), has emerged as a powerful tool for measuring distances to isolated clouds. This approach was successfully applied by \citet{Eswaraiah2013} to LDN 1570, \citet{Barman2015} to CB4, and \citet{Das2015} to six Bok globules.

While these NIR photometric techniques have proven successful for several isolated clouds, their application remains limited, leaving many Bok globules with poorly constrained or entirely unknown distances. Despite their significance as pristine laboratories of low-mass star formation, many Bok globules still lack precise distance estimates, limiting accurate determinations of their physical properties. Building on the methodology of \citet{Choudhury2019}, we present improved distances for five Bok globules---CB4, CB24, CB56, CB60, and CB188. The work pursues three primary goals: (1) to exploit the astrometric accuracy of \textit{Gaia} DR3 to refine distance estimates and resolve literature discrepancies by factors of up to $\sim 3.5$; (2) to provide the first reliable distance measurement for CB56; and (3) to introduce a dual-metric framework that combines weighted-mean and median-based distance estimations, offering enhanced statistical robustness and resistance to outliers for sparsely sampled fields ($N = 16\text{--}103$ stars).

Our analysis employs three layers of validation: internal consistency checks (showing $<10\%$ disagreement between the two estimation methods for all clouds), extinction-break detection to pinpoint the physical cloud location, and independent cross-verification using the \texttt{Bayestar19} 3D dust extinction maps of \citet{Green2019} for CB4 and CB188. These refined measurements have substantial implications, including a revised distance upward by a factor of $\sim 3.5$ for CB188 and establishing the first robust distance baseline for future studies of CB56.

\section{Object Identification}
\label{sec:objects}

We have selected five Bok globules spanning a range of cloud properties and evolutionary stages. Brief descriptions are provided below; their comprehensive observational properties are listed in Table~\ref{tab:cloud_properties}.

\subsection{Individual Cloud Descriptions}

\textbf{CB4} is a small, spherical, and relatively isolated dark globule with an ambiguous distance in the literature ($450$--$600$ pc; \citet{Dickman1983}, \citet{Launhardt2010}, \citet{Barman2015}). Despite its compact appearance, low temperature ($\sim 14$ K), and mass $\sim 1.6 M_{\odot}$ \citep{Launhardt2013}, CB4 lacks embedded sources brighter than the IRAS 100 $\mu$m limit and exhibits cirrus-like characteristics. Its well-ordered magnetic field (position angle $71^{\circ}$; \citet{Sen2005}) appears aligned with the Galactic field, making it a quiescent but scientifically valuable target.

\textbf{CB24} is a small, spherical Bok globule detected at $l = 155.76^{\circ}$, $b = 5.9^{\circ}$ with CO $J = 1 \to 0$ velocity $V_{\text{LSR}} = 4.6$ km s$^{-1}$ \citep{Peterson1998} and a distance of 293 $\pm$ 54 pc from our prior study \citep{Das2015}. Unlike many globules, CB24 lacks associated IRAS sources and exhibits exceptionally low column densities, likely representing a genuine starless core in an early evolutionary stage characterized by minimal star formation activity. No embedded protostars have been detected toward this cloud.

\textbf{CB56} is an asymmetrical, dense cloud at $l = 237.93^{\circ}$, $b = -6.46^{\circ}$ with a major-axis position angle of $170^{\circ}$, for which no distance estimate has been previously published. Two IRAS point sources (07125–2503 and 07125–2507) are associated with it, suggesting ongoing or recent star formation activity \citep{Clemens1988}. The cloud exhibits well-aligned polarization vectors ($p_{\text{av}} = 1.08\%$, $\theta_{\text{av}} = 150.94^{\circ}$) \citep{Chak2014}, indicative of organized magnetic fields and grain alignment, making it an interesting target for studying the relationship between magnetic fields and cloud fragmentation.

\textbf{CB60} is a B-type cloud with an anomalously warm temperature and three associated IRAS point sources (08026--3122, 08029--3118, 08022--3155). Its polarization vectors show a local magnetic-field alignment distinct from the Galactic plane ($p_{\text{av}} = 1.3\%$, $\theta_{\text{av}} = 155.15^{\circ}$) \citep{Chak2014}.

\textbf{CB188} is a low-mass, spherical Bok globule in Aquila at a distance of approximately 260 pc \citep{Das2015}, associated with the Lindblad loop. It harbors a single Class I protostar (IRAS 19179+1129) and represents a gravitationally unstable core with a total mass of $\sim 7$ $M_{\odot}$, temperature $\sim$ 18K, and a central density of $\sim 1 \times 10^5$ cm$^{-3}$ \citep{Kandori2005}. Its Local Standard of Rest velocity ($\sim 7$ km s$^{-1}$) is consistent with that of the surrounding molecular complexes.

\begin{table}
\caption{Physical and Observational Properties of Selected Bok Globules}
\label{tab:cloud_properties}
\centering
\begin{tabular}{lccccc}
\hline
Property & CB4 & CB24 & CB56 & CB60 & CB188 \\
\hline
\multicolumn{6}{c}{\textbf{Positions and Coordinates}} \\
RA (J2000) & 00:39:03 & 04:58:30 & 07:14:36 & 08:04:36 & 19:20:17 \\
Dec (J2000) & +52:51:29 & +52:15:41 & --25:08:54 & --31:30:47 & +11:36:12 \\
Galactic $l$ (deg) & 121.03 & 155.76 & 237.93 & 248.89 & 46.53 \\
Galactic $b$ (deg) & --9.97 & 5.91 & --6.46 & --0.01 & --1.02 \\
\multicolumn{6}{c}{\textbf{Cloud Properties}} \\
Cloud Type & Quiescent & Starless & Dense & B-type, Active & Low-mass, Active \\
IRAS Sources & 1 source & None & 2 sources & 3 sources & 2 sources \\
Temperature & $\sim 14$ K & \ldots & \ldots & Warm & 18 K \\
\multicolumn{6}{c}{\textbf{Magnetic Field Properties}} \\
Mag. Field Angle (deg) & 71 (ord.) & \ldots & 150.94 & 155.15 & 98.5 \\
Polarization $p_{\text{av}}$ (\%) & \ldots & \ldots & $1.08\pm0.56$ & $1.30\pm0.94$ & $3.11 \pm 0.28$$^e$  \\
\multicolumn{6}{c}{\textbf{Physical Parameters}} \\
Mass ($M_{\odot}$) & 1.6$^a$ & \ldots & \ldots & \ldots & $\sim$7$^d$ \\
$V_{\text{LSR}}$ (km s$^{-1}$) & --12$^a$ & 4.6$^b$ & 14.5$^c$ & 13.9$^c$ & $\sim$7$^d$ \\
\multicolumn{6}{c}{\textbf{Literature Distance}} \\
Distance (pc) & 459$\pm$85 & 293$\pm$54 & \ldots & 1500 & 262$\pm$49 \\
Dist. Reference & BD(2015) & D(2015) & \ldots & L(1997) & D(2015) \\
\hline
\end{tabular}
\begin{flushleft}
\small
Ellipses (\ldots) indicate data not available in the literature. BD (2015), L(1997) and D(2015) correspond to \citet{Barman2015}, \citet{Launhardt1997}, and \citet{Das2015}, respectively. 

$^a$\cite{Launhardt2013},
$^b$\cite{Peterson1998},
$^c$\cite{Clemens1991},
$^d$\cite{Kandori2005},
$^e$\cite{Choudhury2022}
\end{flushleft}
\end{table}

\section{Data and Methods}
\label{sec:methods}

\subsection{Near-Infrared Photometric Technique}

\subsubsection{Principle and Historical Development}
The NIR photometric technique exploits wavelength-dependent
extinction by interstellar dust to derive stellar parameters and
extinction values. By comparing observed NIR colours of stars with
their intrinsic main-sequence colours, we derive visual extinction
($A_V$) along each line of sight; sharp increases in $A_V$ with
distance mark cloud locations. \citet{Maheswar2010} developed this
approach using 2MASS $(J-H)$ and $(H-K_S)$ colour indices of
main-sequence stars (A0--K7), which was subsequently refined by
\citet{Choudhury2019} through integration with \textit{Gaia}
parallax data. Our distance estimation procedure follows this same
NIR extinction technique, from which cloud distances are extracted
from extinction--distance profiles and subsequently validated using
the Bayestar19 3D dust map \citep{Green2019}.

\subsubsection{Mathematical Framework}

\textbf{Step 1: Distance from \textit{Gaia} DR3 Parallax}

Stellar distances are obtained directly from \textit{Gaia} DR3 parallaxes:
\begin{equation}
d~(\text{pc}) = \frac{1000}{\pi}
\label{eq:distance}
\end{equation}
where $\pi$ is the parallax in milliarcseconds (\textit{mas}). We note that the \textit{Gaia} EDR3/DR3 astrometry carries a known parallax zero-point offset of approximately $-17~\mu\mathrm{as}$, as determined from quasars \citep{Lindegren2021}. Since \textit{Gaia} DR3 astrometry is identical to EDR3, this characterisation applies directly to the data used in this study. This zero-point offset was not corrected for explicitly; however, at the parallax values relevant to this study ($\sim0.7$--$3.3$ mas), corresponding to distances of $\sim300$--$1500$ pc, the offset introduces a systematic distance uncertainty of $\sim0.5$--$2.5\%$. This systematic falls comfortably within the physically meaningful distance uncertainties of this study, characterised by $\mathrm{IQR}/(2d_{\mathrm{median}})$ and ranging from $\sim3$--$9\%$ across the five clouds (Table~\ref{tab:median_results}). Neglecting the zero-point correction causes the inferred distances to be slightly overestimated; however, the magnitude of this bias remains smaller than the distance spreads characterised by the IQR-based uncertainty measure, and therefore does not materially affect our distance estimates or conclusions.

\textbf{Step 2: Observed Colors from 2MASS}

For each star, observed magnitudes ($J$, $H$, $K_S$) from 2MASS yield observed colors:
\begin{equation}
(J-H)_{\text{obs}} = m_{J} - m_{H}
\label{eq:jh_obs}
\end{equation}
\begin{equation}
(H-K_S)_{\text{obs}} = m_{H} - m_{K_S}
\label{eq:hk_obs}
\end{equation}

\textbf{Step 3: Iterative Extinction and Spectral Type Determination}

We iteratively determine extinction $A_V$ and spectral type through the following procedure:

\begin{enumerate}[(a)]
\item Assume trial extinction value $A_V$ (ranging from 0.1 to 20 mag in 0.1 mag steps)
\item Apply \citet{Rieke1985} reddening law to compute extinction in each band:
\begin{align}
A_J &= 0.282 \times A_V \label{eq:aj}\\
A_H &= 0.175 \times A_V \label{eq:ah}\\
A_{K_S} &= 0.112 \times A_V \label{eq:aks}
\end{align}
\item Calculate dereddened (intrinsic) colors:
\begin{align}
(J - H)_0 &= (J - H)_{\text{obs}} - (A_J - A_H) = (J-H)_{\text{obs}} - 0.107 \times A_V \label{eq:jh0}\\
(H - K_S)_0 &= (H - K_S)_{\text{obs}} - (A_H - A_{K_S}) = (H-K_S)_{\text{obs}} - 0.063 \times A_V \label{eq:hk0}
\end{align}
\item Determine spectral type by minimizing $\chi^2$ difference between observed dereddened colors and intrinsic main-sequence colors (A0–K7) from \citet{Cox2000} calibrations:
\begin{equation}
\chi^2 = \frac{[(J-H)_{0,\text{obs}} - (J-H)_{0,\text{MS}}]^2}{\sigma^2_{J-H}} + \frac{[(H-K_S)_{0,\text{obs}} - (H-K_S)_{0,\text{MS}}]^2}{\sigma^2_{H-K_S}}
\label{eq:chi2}
\end{equation}
\item For the best-fit spectral type, obtain absolute magnitude $M_{K_S}$ from main-sequence relations.
\item Compare calculated distance $d_{\text{calc}}$ from the extinction-corrected distance modulus:
\begin{equation}
d_{\text{calc}} = 10^{(m_{K_S} - A_{K_S} - M_{K_S} + 5)/5}~\text{pc}
\label{eq:dcalc}
\end{equation}
where $A_{K_S} = 0.112 \times A_V$ (Eq.~\ref{eq:aks}). Compare $d_{\text{calc}}$ with \textit{Gaia} distance $d_{\text{Gaia}}$. Iterate $A_V$ until convergence ($|d_{\text{calc}} - d_{\text{Gaia}}|$ minimized).
\end{enumerate}

\textbf{Step 4: Final Extinction Value}

The converged $A_V$ represents the total visual extinction along the line of sight to each star at its \textit{Gaia}-determined distance.

\subsubsection{Critical Distinction: NIR Photometry vs Optical Astrometry}
\label{sec:nir_vs_optical}

\textbf{Important Methodological Note:} This technique employs a dual-data approach that is critical to its success in studying moderately extincted clouds. We use \textit{near-infrared} (NIR) photometry from 2MASS ($J = 1.25~\mu$m, $H = 1.65~\mu$m, $K_s = 2.17~\mu$m) to measure extinction \textit{through} the clouds, not optical photometry. NIR wavelengths experience significantly lower extinction than optical wavelengths, with the extinction coefficients from the \citet{Rieke1985} reddening law (Eqs. ~\ref{eq:aj}--\ref{eq:aks}) giving $A_J/A_V = 0.282, A_H/A_V = 0.175$, and $A_{K_s}/A_V = 0.112$ — factors of approximately 3.5, 5.7, and 9 times lower than visual extinction, respectively. This means a cloud with $A_V = 20$ mag (optically opaque) has only $A_J \approx 5.6$ mag, $A_H \approx 3.5$ mag, and $A_{K_s} \approx 2.2$ mag in the NIR bands—well within 2MASS detection limits for background stars.

\textit{Gaia} DR3 parallaxes are used only for measuring distances to field stars via trigonometric parallax (Eq.~\ref{eq:distance}), not for obtaining photometry through the cloud itself. This separation of functions—NIR photometry for extinction measurement + optical astrometry for distance determination—is what enables the method to work effectively for clouds where optical photometry would be severely compromised by extinction.

The method is most reliable for clouds with $A_{V,\text{max}} \lesssim 15$ mag---the general upper limit of 2MASS-based NIR extinction techniques beyond which even the $J$-band becomes too extincted for sufficient background star detection. In practice, the operative range of the present technique is further restricted to $A_V \lesssim 4$ mag by the $J - K_S \leq 0.75$ colour criterion (Section~\ref{sec:method_limits}), which excludes stars reddened beyond this level to avoid contamination by unreddened M-type dwarfs \citep[see Fig.~1 of][]{Maheswar2010}. All five Bok globules in this study satisfy $A_{V,\text{max}} \lesssim 4$ mag, confirming the technique is applicable to the present sample.

\subsubsection{Data Selection Criteria}
\label{sec:method_limits}

We obtained near-infrared photometry for stars within a $30' \times 30'$ field around each globule from the 2MASS Point Source Catalog \citep{Cutri2003}. Selected sources met the following criteria:

\begin{enumerate}[(a)]
\item \textbf{Quality flags}: Quality flag (Qflag) = `AAA' in all three filters ($J$, $H$, $K_S$), indicating high signal-to-noise ratio (SNR $> 10$) and reliable detections.
\item \textbf{Photometric uncertainties}: Uncertainties $\le 0.045$ mag in all three filters, ensuring accurate color indices.
\item \textbf{Spectral type range}: $J - K_S \le 0.75$ to exclude unreddened M-type stars. This criterion is critical because unreddened M-type dwarfs occupy the same reddening vector as reddened A0–K7 dwarfs in the $(J - H, H - K_S)$ color diagram \citep{Maheswar2010}, potentially introducing systematic biases in extinction determination. By restricting to $J - K_S \le 0.75$, we exclude the vast majority of unreddened M-type stars while retaining reddened early- to mid-type stars most sensitive to extinction variations.
\end{enumerate}

Parallax data were obtained from the \textit{Gaia} DR3 catalog \citep{Gaia2016, Gaia2022} with the requirement that parallax $\pi > 0$ (physically meaningful distances).

\subsection{Data Processing Pipeline}

Our analysis follows a systematic procedure:

\begin{enumerate}
\item \textbf{Source Selection}: Extract all 2MASS sources in the $30' \times 30'$ field meeting quality criteria (a)-(c) from Section~\ref{sec:methods}, subsection~\ref{sec:method_limits}.
\item \textbf{Parallax Cross-Matching}: Match 2MASS sources with \textit{Gaia} DR3 parallaxes within $1''$ angular separation; calculate distances $d~(\text{pc}) = 1000/\pi$ (parallax in milliarcseconds).
\item \textbf{Extinction Calculation}: For each star, calculate visual extinction $A_V$ by iteratively fitting trial values (0.1 to 20 mag in 0.1 mag steps), dereddening colors, determining best-fit spectral type by minimizing $\chi^2$ between observed dereddened colors and main-sequence standards, and converging when photometric distance matches \textit{Gaia} distance.
\item \textbf{Distance Determination}: Apply the dual-metric approach combining weighted mean calculation with median distance estimation, as detailed in Section~\ref{sec:distance_method}.
\item \textbf{Validation}: Cross-match results with \textit{Gaia} DR3 uncertainties to ensure parallax quality. Verify agreement between weighted mean and median distances (target: $<$10\% difference; see Section~\ref{sec:validation} for justification) and confirm consistent extinction signatures indicative of genuine cloud structure. 
\end{enumerate}

\subsection{Extinction Uncertainty Propagation}

The uncertainty in $A_V$ for each star arises from multiple sources: photometric measurement errors in 2MASS colors, uncertainties in spectral type determination, and systematic uncertainties in the reddening law coefficients. We propagate these uncertainties through the dereddening process to obtain the total $A_V$ uncertainty for each star.

The photometric uncertainties from 2MASS ($\sigma_{J-H}$ and $\sigma_{H-K_S}$) propagate through the color-extinction relations with different sensitivities in each color index. The spectral type fitting procedure introduces additional uncertainty, typically equivalent to $\sim 0.5$--1 spectral subclass uncertainty, which translates to approximately $0.1$--$0.2$ mag in $A_V$. Furthermore, systematic uncertainties in the \citet{Rieke1985} reddening law coefficients contribute approximately 5--10\% of the derived $A_V$ value.

These individual uncertainty components are combined in quadrature to yield the total extinction uncertainty $\sigma_{A_V}$ for each star. This comprehensive error budget ensures that systematic errors in spectral type determination or reddening law assumptions do not dominate the extinction measurement uncertainty.

For distance determination, we apply a two-tier selection threshold based on extinction significance:

\begin{itemize}
\item \textbf{Standard threshold}: $A_V - \sigma_{A_V} > 0.4$ mag, corresponding to statistically significant extinction detections. This threshold is applied to CB24, CB56, CB60, and CB188.
\item \textbf{Relaxed threshold (low-extinction clouds only)}: $A_V - \sigma_{A_V} > 0.2$ mag for clouds where no stars meet the strict criterion, preventing complete sample elimination while maintaining reasonable detection confidence. This relaxed criterion is applied only to CB4 due to its low optical depth and substantial foreground extinction.
\end{itemize}

This approach ensures that only stars with robust, statistically significant extinction measurements contribute to the cloud distance determination, minimizing contamination from foreground stars or measurement artifacts.

\subsection{Sample Selection and Cloud Morphology}

\begin{figure}
\centering
\includegraphics[width=\textwidth]{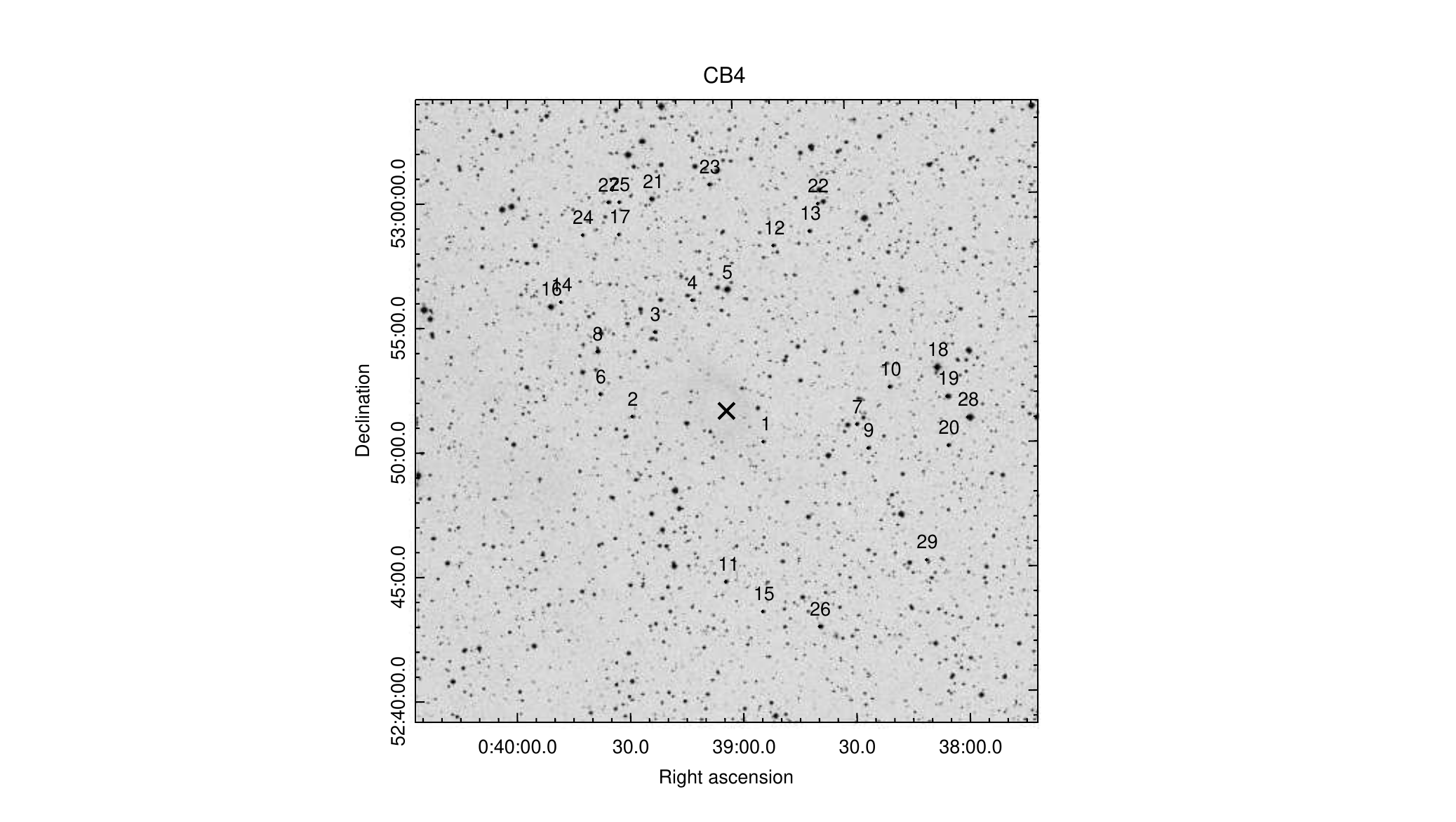}
\caption{DSS R-band image of Bok Globule CB4 showing the 29 selected 2MASS
field stars used in the analysis, chosen according to the criteria described in
Section~\ref{sec:methods} (subsection~3.1.4).}
\label{fig:cb4}
\end{figure}

\begin{figure}
\centering
\includegraphics[width=\textwidth]{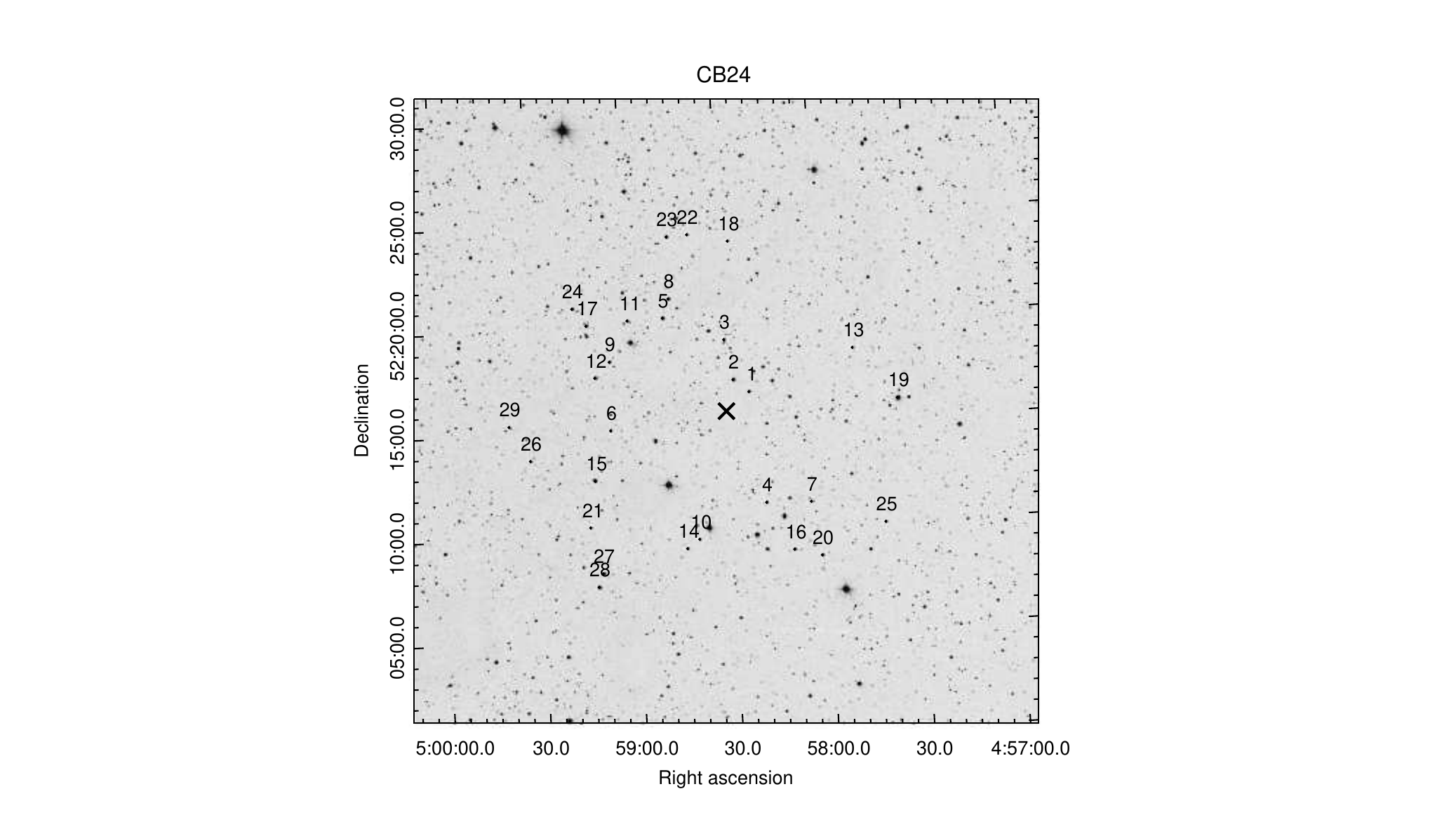}
\caption{Same as Figure~\ref{fig:cb4} but for CB24 (29 selected stars).}
\label{fig:cb24}
\end{figure}

\begin{figure}
\centering
\includegraphics[width=\textwidth]{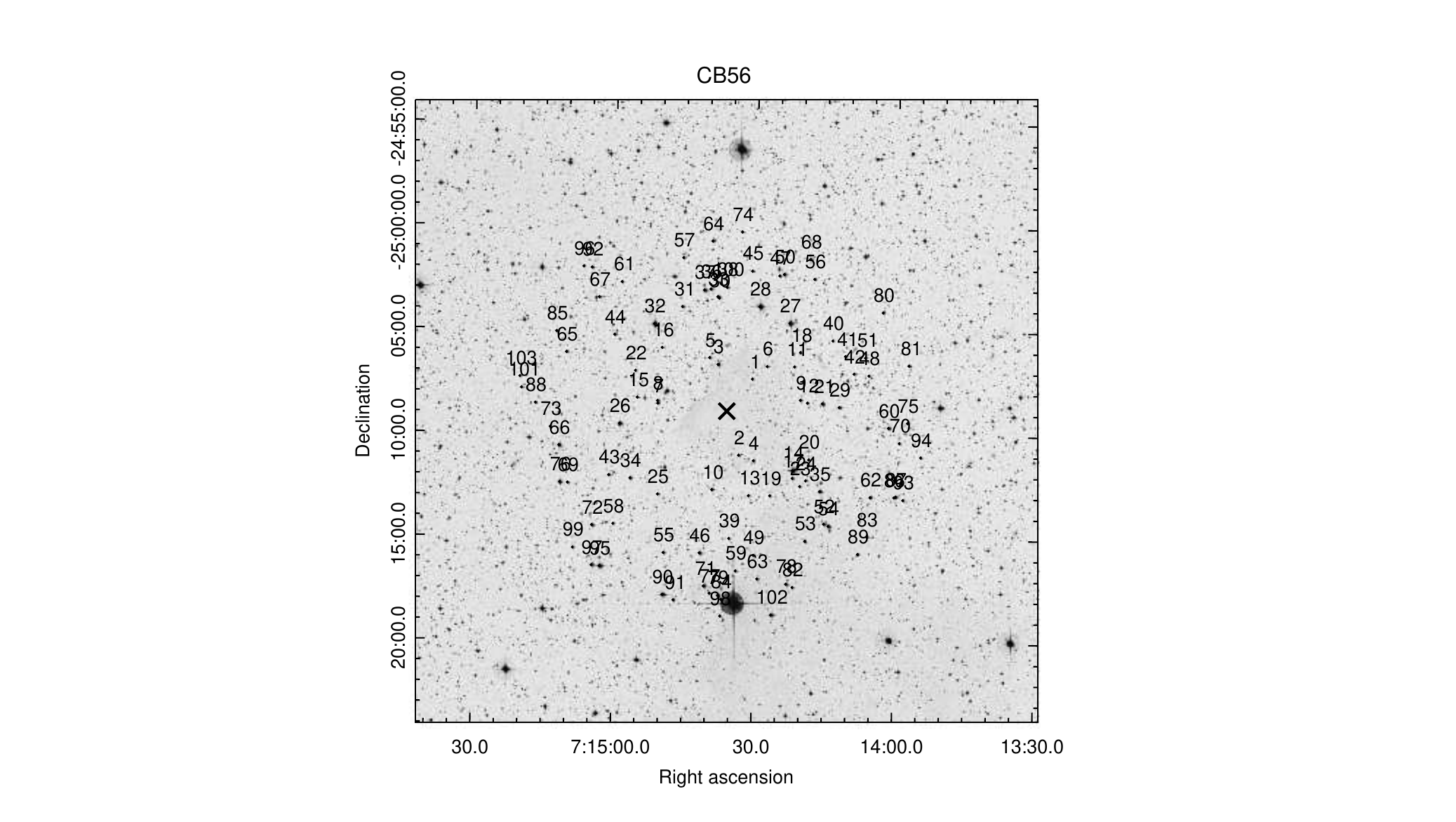}
\caption{Same as Figure~\ref{fig:cb4} but for CB56 (103 selected stars).}
\label{fig:cb56}
\end{figure}

\begin{figure}
\centering
\includegraphics[width=\textwidth]{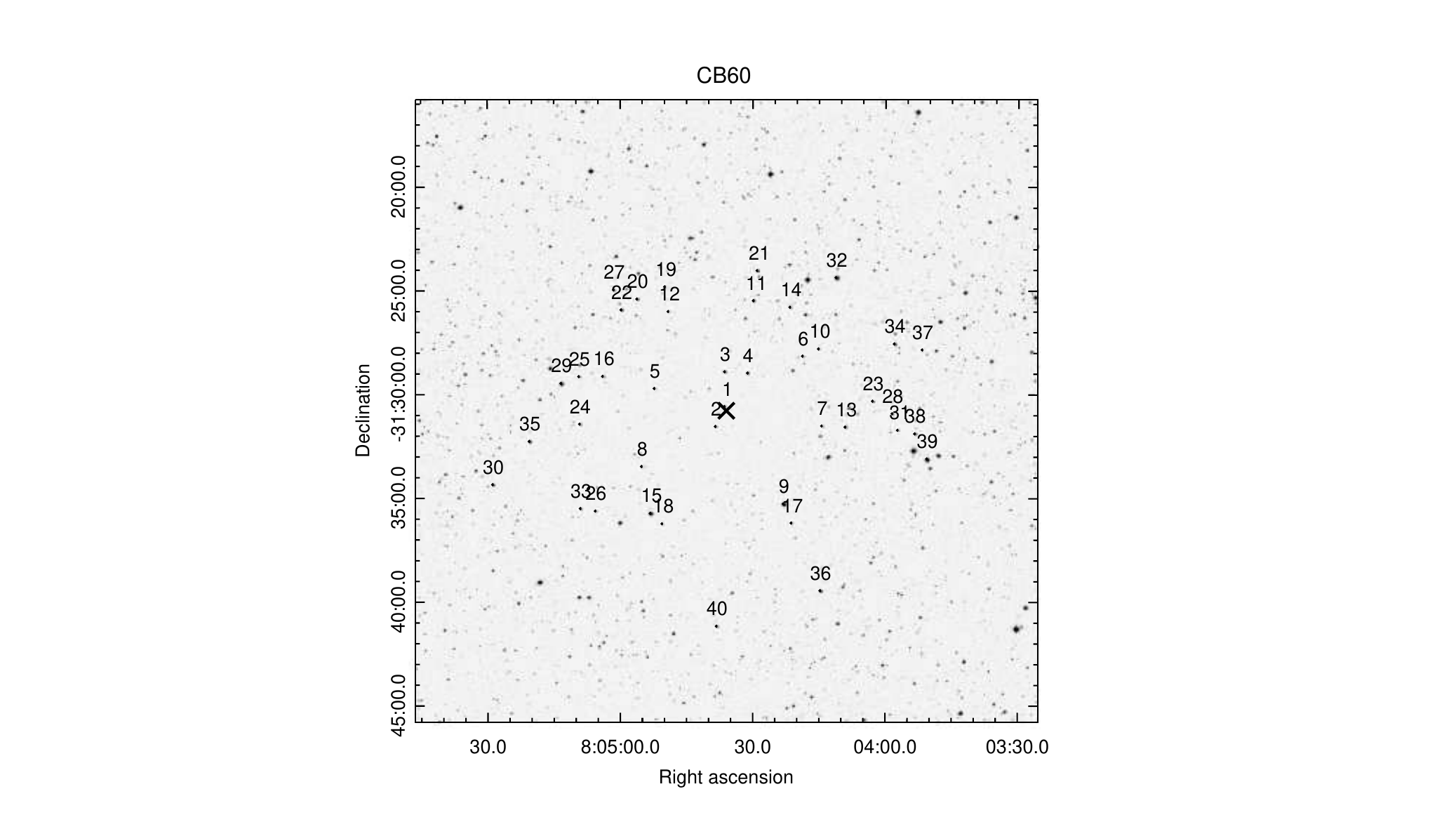}
\caption{Same as Figure~\ref{fig:cb4} but for CB60 (40 selected stars).}
\label{fig:cb60}
\end{figure}

\begin{figure}
\centering
\includegraphics[width=\textwidth]{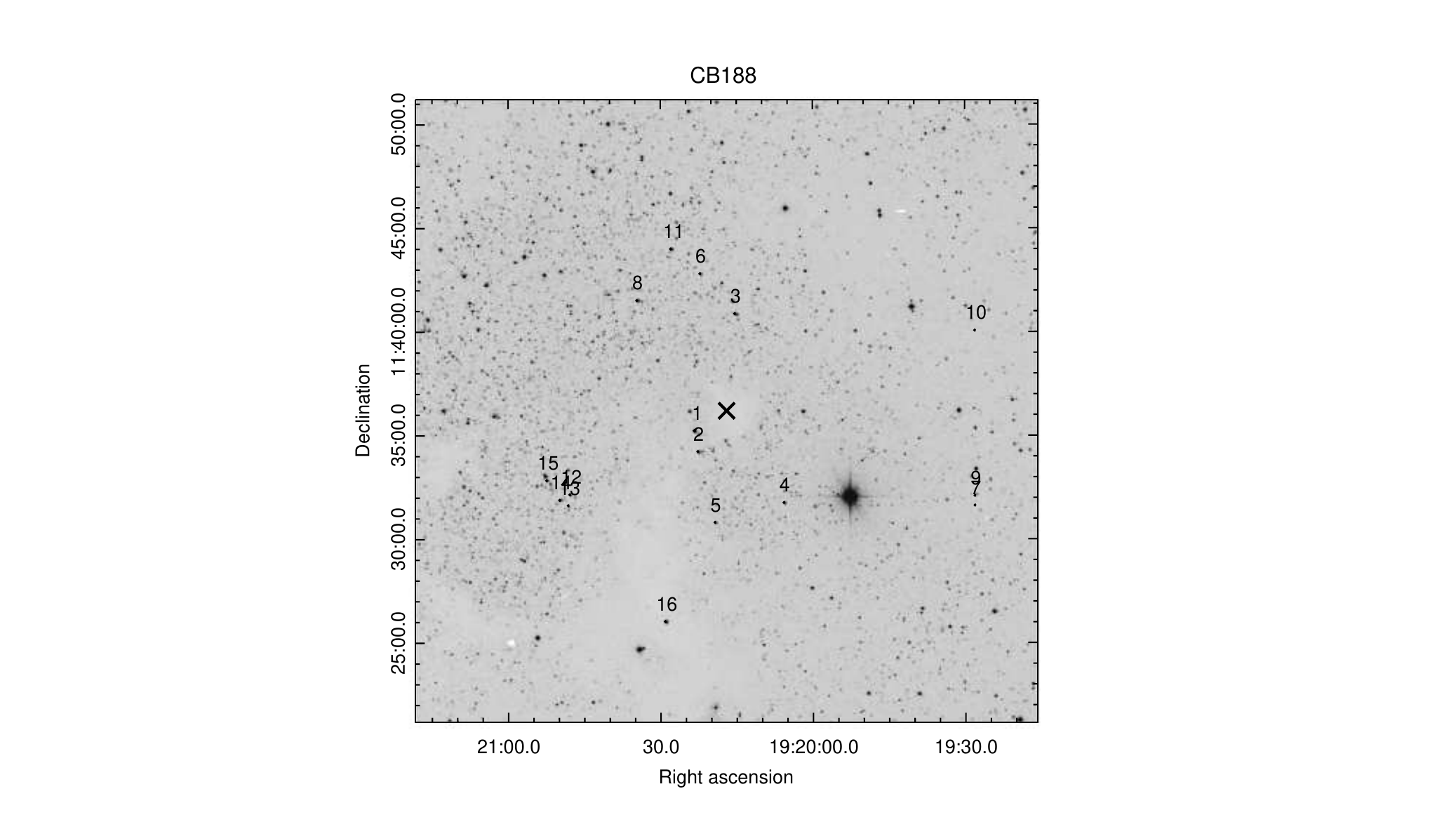}
\caption{Same as Figure~\ref{fig:cb4} but for CB188 (16 selected stars).}
\label{fig:cb188}
\end{figure}

The five Bok globules analyzed in this study are shown in Figures~\ref{fig:cb4}--\ref{fig:cb188}, where each panel displays a Digital Sky Survey (DSS) R-band image of the respective globule along with the selected 2MASS field stars used in the distance determination. The DSS images clearly reveal the morphology and projected structure of each cloud, providing visual confirmation of their compact, isolated nature. The field stars shown in these figures were selected following the comprehensive quality criteria specified in Section~\ref{sec:methods}, subsection 3.1.3.

\textbf{CB4} (Figure~\ref{fig:cb4}, 29 selected stars): The globule appears as a dark, compact, nearly spherical structure with clear boundaries against the background stellar field.

\textbf{CB24} (Figure~\ref{fig:cb24}, 29 selected stars): This globule exhibits similar spherical morphology to CB4, but with slightly less contrast against the background. The selected stars provide good spatial coverage for extinction analysis.

\textbf{CB56} (Figure~\ref{fig:cb56}, 103 selected stars): Unlike CB4 and CB24, CB56 exhibits asymmetrical morphology with an irregular boundary, consistent with its dense, structured appearance. The substantial number of available stars enables robust statistical analysis.

\textbf{CB60} (Figure~\ref{fig:cb60}, 40 selected stars): This globule appears compact but exhibits patchy structure characteristic of B-type clouds with active star formation.

\textbf{CB188} (Figure~\ref{fig:cb188}, 16 selected stars): Despite the smallest sample size, this number is adequate for distance determination. The cloud's spherical morphology and compact structure are clearly evident.

\section{Distance Estimation Method}
\label{sec:distance_method}

\subsection{Data Sources and Extinction Measurements}

Stellar distances and visual extinctions $A_V$ have been computed as described in
Section~\ref{sec:methods}. To focus on stars in the immediate vicinity of each
cloud, we have defined a maximum projected radial distance $r_{\text{max}}$ from the
cloud center, excluding stars projected beyond this radius. The $r_{\text{max}}$
values for each cloud (Table~\ref{tab:median_results}) were chosen to encompass
the main cloud structure visible in DSS images while excluding unrelated field
stars. Specifically, $r_{\mathrm{max}}$ was determined through visual inspection of each DSS R-band image (Figures~\ref{fig:cb4}--\ref{fig:cb188}) by examining the surrounding stellar field and extinction morphology radially outward from the cloud center. The adopted $r_{\mathrm{max}}$ values define the maximum projected radial extent of the localized stellar sampling region used for the extinction analysis. These radii were selected such that sufficient foreground and background stars could be included for constructing reliable extinction--distance profiles while minimizing contamination from unrelated large-scale field structure. The extinction--distance profiles (Figures~\ref{fig:cb4_ext}--\ref{fig:cb188_ext}) further verify that the selected regions exhibit coherent extinction rises associated with the clouds rather than scattered distributions characteristic of unrelated field stars. We emphasise that these sampling radii (6.0--7.0 arcmin) are operational parameters chosen for the stellar extinction analysis and do not correspond to the intrinsic angular diameters of the opaque cloud cores as catalogued by \citet{Clemens1988}. The catalogue dimensions describe the visually darkest core regions; extinction-distance analysis requires sampling of the surrounding stellar field to identify foreground and background populations straddling the cloud, necessitating a larger effective radius.

\subsection{Selection Criteria}

To ensure that distance estimates are based on stars with reliable extinction measurements and well-determined parallaxes, we applied the following selection criteria:

\subsubsection{Standard Selection Criteria (CB24, CB56, CB60, CB188)}

\begin{enumerate}[(i)]
\item $A_V - \sigma_{A_V} > 0.4$: This requires that the net extinction — after subtracting the 1-sigma uncertainty — still exceeds 0.4 mag, ensuring a statistically significant and physically meaningful extinction detection above the foreground noise floor. This threshold preferentially selects moderately to heavily reddened stars where extinction determination is most reliable.
    
\item $d/\sigma_d > 3$: Distance measurement uncertainty should be $\le 33\%$ of the distance itself. For \textit{Gaia} DR3 parallaxes, this typically restricts the sample to $V < 14$ mag or parallax uncertainties $\sigma_{\pi} < 0.3$ \textit{mas}, ensuring reliable trigonometric distances.
    
\item $A_V/\sigma_{A_V} > 1.5$: Extinction-to-noise ratio must exceed 1.5, providing an additional buffer against systematic errors in spectral type determination or dust model assumptions.
\end{enumerate}

\subsubsection{Relaxed Criteria (CB4: $A_V - \sigma_{A_V} > 0.2$, $d/\sigma_d > 3$, $A_V/\sigma_{A_V} > 1.2$)}
Application of standard criteria to CB4 yielded only 8 candidate stars, which is at the lower limit for reliable distance determination via weighted
averaging. Analysis of the extinction-distance plot for CB4 revealed that the cloud's relatively low optical depth ($A_{V,\mathrm{max}} \sim 3$--4 mag)
combined with its substantial foreground column density makes stars meeting stringent criteria rare. Consequently, we relaxed the net-extinction
floor to $A_V - \sigma_{A_V} > 0.2$ mag, requiring the extinction to remain positive and physically meaningful after subtracting the 1-sigma uncertainty,
retaining $A_V/\sigma_{A_V} > 1.2$ for reasonable signal-to-noise reliability. This adjustment increased the usable sample to 29 stars,
enabling robust weighted averaging. We note that while CB4 distances should be interpreted with slightly higher uncertainty due to the relaxed
criteria, the results remain physically meaningful and comparable to independent estimates (Section~\ref{sec:comparison_3d}).

\subsection{Distance Determination Procedure}

Our distance determination employs a systematic five-step approach that combines weighted mean calculation with median distance estimation and extinction break analysis.

\subsubsection{Step 1: Weighted Mean Distance (Reference Point)}

For each cloud, we calculated the weighted mean distance using inverse-variance weighting:
\begin{equation}
\langle d \rangle = \frac{\sum_i w_i d_i}{\sum_i w_i}
\label{eq:weighted_mean}
\end{equation}
\begin{equation}
w_i = \frac{1}{\sigma_{d,i}^2}
\label{eq:weights}
\end{equation}
\begin{equation}
\sigma_{\langle d \rangle} = \frac{1}{\sqrt{\sum_i w_i}}
\label{eq:sigma_weighted}
\end{equation}
where $d_i$ is the distance of the $i$-th star, $w_i$ are weights inversely proportional to squared distance uncertainties, and $\sigma_{d,i}$ is the uncertainty in stellar distance. This weighting scheme emphasizes high-precision measurements and provides formal uncertainty estimates. The weighted mean $\langle d \rangle$ serves as an important reference point for subsequent extinction break analysis.

\subsubsection{Step 2: Extinction Break Analysis and Distance Limits}

The visual extinction ($A_V$) is plotted as a function of heliocentric distance for all selected stars (Figures~\ref{fig:cb4_ext}--\ref{fig:cb188_ext}). A characteristic sharp transition in extinction marks the cloud location. The analysis proceeds as follows:

\begin{enumerate}
\item \textbf{Reference Point}: Use $\langle d \rangle$ as the reference heliocentric distance.
\item \textbf{Extinction Distribution Analysis}: Examine the distribution of extinction values relative to $\langle d \rangle$:
\begin{itemize}
\item Stars at distances $< \langle d \rangle$ typically show lower extinction (foreground ISM)
\item Stars at distances near/within the cloud show sharp extinction increase
\item Stars at distances $> \langle d \rangle$ show high extinction or plateau
\end{itemize}
\item \textbf{Identify Extinction Break Location}: The extinction break is identified as the distance at which the first sharp increase 
in $A_V$ above the foreground level occurs, indicating the cloud's front edge \citep{Maheswar2010, Eswaraiah2013}. Operationally, the 
foreground extinction level is estimated as the mean $A_V$ of all selected stars with $d < \langle d \rangle$, and the break is 
identified at the shortest distance where $A_V$ exceeds this foreground level by more than $1\sigma_{A_V}$. The transition zone 
boundaries ($d_\mathrm{lower}$, $d_\mathrm{upper}$) are then defined as the distances of the last foreground-level star and the first 
significantly reddened star on either side of this break, respectively. This break location may occur before, at, or after $\langle d \rangle$.

\item \textbf{Establish Distance Limits}: Based on extinction distribution:
\begin{itemize}
\item \textbf{Case A (Extinction break $<$ weighted mean)}: High-extinction stars found primarily at $d < \langle d \rangle$
\begin{itemize}
\item Lower limit: $d_{\text{lower}}$ = extinction break distance
\item Upper limit: $d_{\text{upper}} = \langle d \rangle$
\end{itemize}
\item \textbf{Case B (Extinction break $>$ weighted mean)}: High-extinction stars found primarily at $d > \langle d \rangle$
\begin{itemize}
\item Lower limit: $d_{\text{lower}} = \langle d \rangle$
\item Upper limit: $d_{\text{upper}}$ = extinction break distance
\end{itemize}
\end{itemize}
\item \textbf{Cloud Distance Range}: The true cloud distance lies between $d_{\text{lower}}$ and $d_{\text{upper}}$, with both limits informed by the extinction-distance relationship.
\end{enumerate}

\subsubsection{Step 3: Median Distance as final reported estimator}

From the population of selected stellar distances, we calculate the
median distance $d_\mathrm{median}$ as the final reported distance estimator
(Eq.~\ref{eq:median}). The median is preferred over the weighted mean
as the final reported result because it is robust against outliers, makes no
assumption of Gaussian statistics (ideal for small samples of
$N=16$--103 stars), and is validated by the weighted mean: agreement
between the two confirms reliability. The weighted mean $\langle d
\rangle$ serves as an important reference point and provides formal
uncertainty estimates for cross-validation.

\begin{equation}
d_{\mathrm{median}} =
\begin{cases}
d_{(N+1)/2} & \text{for odd } N \\[4pt]
\dfrac{d_{N/2}+d_{(N/2+1)}}{2} & \text{for even } N
\end{cases}
\label{eq:median}
\end{equation}

where $N$ is the number of selected stars and $d_{(i)}$ represents
the $i$-th distance in sorted order.

\subsubsection{Step 4: Uncertainty Quantification}

The interquartile range (IQR) provides a non-parametric uncertainty estimate:
\begin{equation}
\text{IQR} = Q_3 - Q_1
\label{eq:iqr}
\end{equation}
where $Q_1$ (first quartile, 25th percentile) and $Q_3$ (third quartile, 75th percentile) represent the distances below and above which 25\% and 75\% of the stellar distances fall, respectively. The IQR quantifies the spread in individual stellar distance measurements and does not depend on formal error propagation.

\subsubsection{Step 5: Validation Through Agreement Assessment}\label{sec:validation}

Agreement between weighted mean and median is quantified as:
\begin{equation}
\text{Agreement}~(\%) = \frac{|\langle d \rangle - d_{\text{median}}|}{d_{\text{median}}} \times 100\%
\label{eq:agreement}
\end{equation}

Agreement $<10\%$ indicates good consistency; $>20\%$ would suggest a
multi-modal distribution requiring investigation. The $<$10\% threshold is chosen as a conservative upper bound on 
the expected sampling scatter between the two estimators for small 
stellar samples ($N = 16$--103): for a unimodal distance distribution, 
the weighted mean and median are expected to be similar within sampling uncertainties, and any 
agreement within 10\% is consistent with sampling noise alone. 
A discrepancy exceeding 20\% would indicate a genuinely multi-modal 
or contaminated stellar distance distribution requiring further 
investigation \citep{Schlafly2014}. All five clouds in this study achieve $\le\textbf{8.0\%}$ agreement, confirming that neither method is systematically
biased.

\section{Results}
\label{sec:results}

\subsection{Distance Determination Sample Properties}

The properties of stars used in the distance determination for each cloud are summarized in Table~\ref{tab:median_results}. The sample sizes for standard criteria ranged from 16 stars (CB188) to 103 stars (CB56), providing statistically robust distance estimates for all clouds. CB4, requiring relaxed criteria as discussed in Section~\ref{sec:distance_method}, was based on 29 stars meeting the modified threshold.

\subsection{Extinction-Distance Analysis and Cloud Location Identification}

The visual extinction ($A_V$) determined for all field stars is plotted as a
function of both projected radial distance from the cloud center and heliocentric
distance in Figures~\ref{fig:cb4_ext}--\ref{fig:cb188_ext}, following the
procedure of Section~\ref{sec:distance_method}. Selected stars meeting stringent
criteria are highlighted in black, while non-selected stars are shown as gray
circles. Yellow shaded regions indicate the probable cloud distance interval(s)
identified from extinction break and 3D dust map comparisons. For all five clouds,
the extinction-distance plots display clear transitions marking cloud locations,
with median distances agreeing well with weighted mean distances ($\le8.0\%$
difference), validating our dual-metric approach.

\subsection{CB4: Extinction Structure and Distance Determination}

Figure~\ref{fig:cb4_ext} shows the extinction profile of CB4. The cloud exhibits confined structure within $r_{\text{max}} = 7.0$ arcmin (panel i), with a sharp extinction rise in the range 763--769 pc (panel ii), where the lower limit marks the extinction break distance and the upper limit corresponds to the weighted mean distance ⟨d⟩ = 769.4 pc (rounded to the nearest parsec in Table~\ref{tab:median_results}). The weighted mean distance ($\langle d \rangle = 769.4 \pm 8.3$ pc) is consistent with the extinction break, indicating excellent convergence between methods. The median distance ($d_{\text{median}} = 766$ pc) falls at the center of this transition zone, with only 0.4\% difference from the weighted mean---the best agreement among all five clouds. The yellow shaded region (763--769 pc) from the 3D dust extinction map of \citet{Green2019} independently confirms our distance determination, providing strong external validation.

\subsection{CB24: Extinction Structure and Distance Determination}

Figure~\ref{fig:cb24_ext} reveals CB24's compact structure. The extinction break occurs at 334--374 pc (yellow shaded region), marginally below the weighted mean distance ($\langle d \rangle = 373.9 \pm 1.5$ pc), indicating asymmetric stellar distribution with more high-extinction stars at smaller distances. The median distance ($d_{\text{median}} = 354$ pc) falls within this transition zone, showing 5.6\% agreement with the weighted mean---well within our validation criterion. This offset demonstrates the median's robustness to asymmetric distributions, as both metrics converge on a consistent distance range despite the non-uniform stellar sampling.

\subsection{CB56, CB60, and CB188: Extinction Structures}

\textbf{CB56} (Figure~\ref{fig:cb56_ext}): This cloud shows asymmetrical structure extending beyond $r_{\text{max}} = 6.0$ arcmin. The extinction rise occurs at 467--499 pc, with weighted mean $\langle d \rangle = 498.6 \pm 3.6$ pc and median $d_{\text{median}} = 483$ pc (3.2\% agreement). The substantial sample (103 stars) enables robust distance determination despite the cloud's irregular morphology.

\textbf{CB60} (Figure~\ref{fig:cb60_ext}): As the most distant cloud, CB60 shows confined structure within $r_{\text{max}} = 7.0$ arcmin but larger distance scatter due to parallax limitations at $\sim 1.2$ kpc. The extinction rise (1124--1162 pc) yields weighted mean $\langle d \rangle = 1162.2 \pm 11.9$ pc and median $d_{\text{median}} = 1143$ pc (1.7\% agreement), demonstrating the method's applicability even at larger distances.

\textbf{CB188} (Figure~\ref{fig:cb188_ext}): This compact cloud exhibits well-defined structure with extinction rise at 852--1000 pc (yellow shaded region from 3D dust map). Weighted mean $\langle d \rangle = 852.0 \pm 8.0$ pc and median $d_{\text{median}} = 926$ pc show 8.0\% agreement, the largest offset among our sample but still within acceptable limits, possibly reflecting the relatively small sample size and broader extinction transition region.

\begin{figure}
\centering
\includegraphics[width=\textwidth]{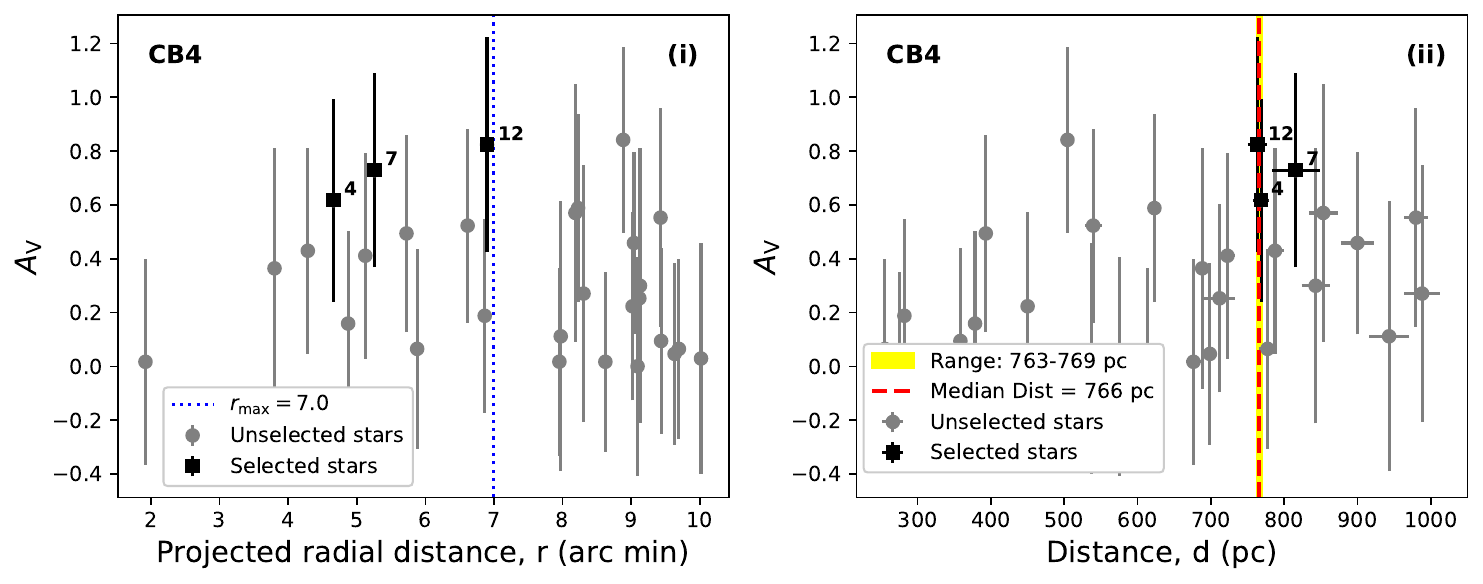}
\caption{Extinction analysis for CB4. \textbf{(i)} Extinction ($A_V$) as a function of projected radial distance from the cloud center. The dotted vertical line marks $r_{\text{max}} = 7.0$ arcmin, indicating the cloud boundary. Selected stars (filled black squares) meeting the selection criteria are marked alongside unselected stars (gray filled circles). \textbf{(ii)} Extinction versus distance for stars within the CB4 region. The yellow shaded area (763--769 pc) indicates the probable cloud location interval. The red dashed line indicates the median distance of 766 pc. Selected stars are highlighted in black, with unselected stars shown as gray circles.}
\label{fig:cb4_ext}
\end{figure}

\begin{figure}
\centering
\includegraphics[width=\textwidth]{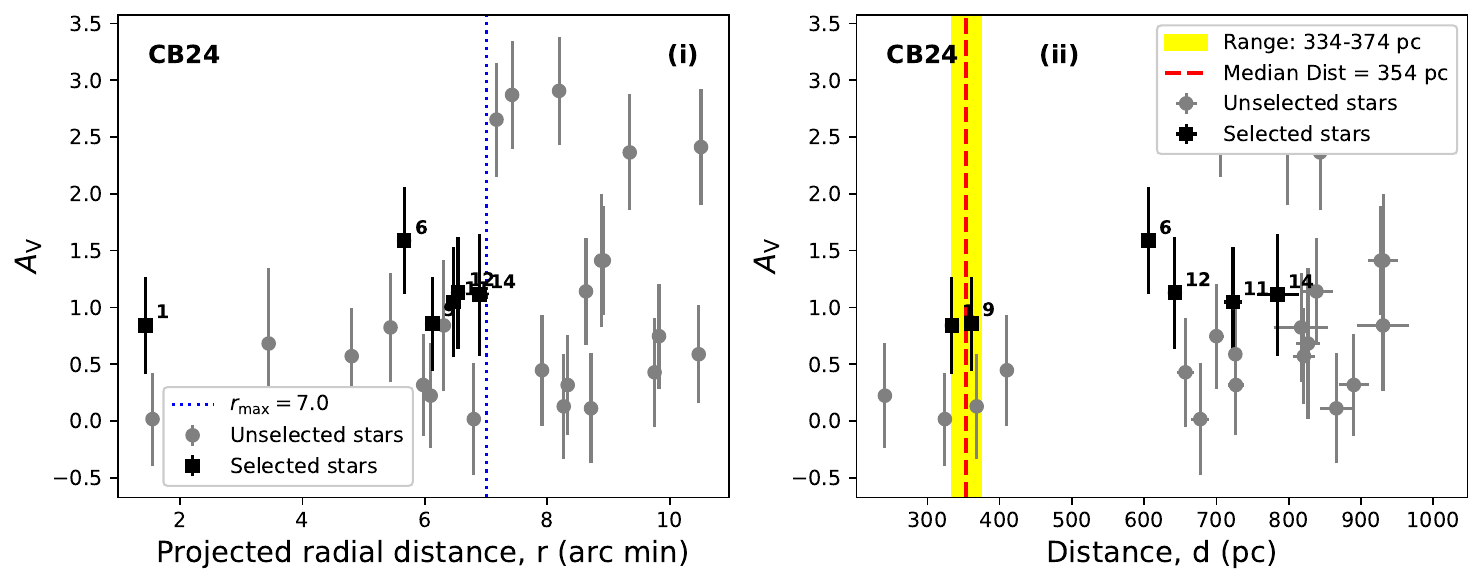}
\caption{Extinction analysis for CB24. \textbf{(i)} Extinction ($A_V$) as a function of projected radial distance from the cloud center. The dotted vertical line marks $r_{\text{max}} = 7.0$ arcmin. \textbf{(ii)} Extinction versus distance for stars within the CB24 region (334--374 pc, yellow shaded area). The red dashed line indicates the median distance of 354 pc. Selected stars are highlighted in black, with unselected stars shown as gray circles.}
\label{fig:cb24_ext}
\end{figure}

\begin{figure}
\centering
\includegraphics[width=\textwidth]{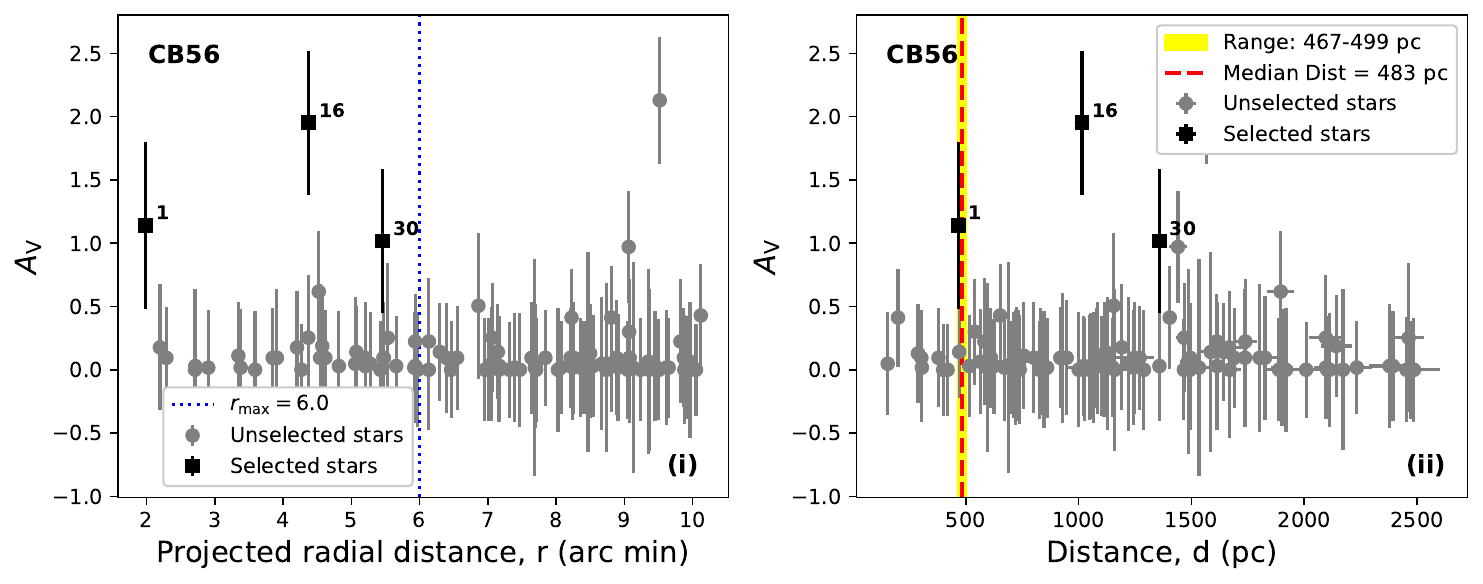}
\caption{Extinction analysis for CB56. \textbf{(i)} Extinction ($A_V$) as a function of projected radial distance from the cloud center. The dotted vertical line marks $r_{\text{max}} = 6.0$ arcmin. \textbf{(ii)} Extinction versus distance for stars within the CB56 region (467--499 pc, yellow shaded area). The red dashed line indicates the median distance of 483 pc. Selected stars are highlighted in black, with unselected stars shown as gray circles.}
\label{fig:cb56_ext}
\end{figure}

\begin{figure}
\centering
\includegraphics[width=\textwidth]{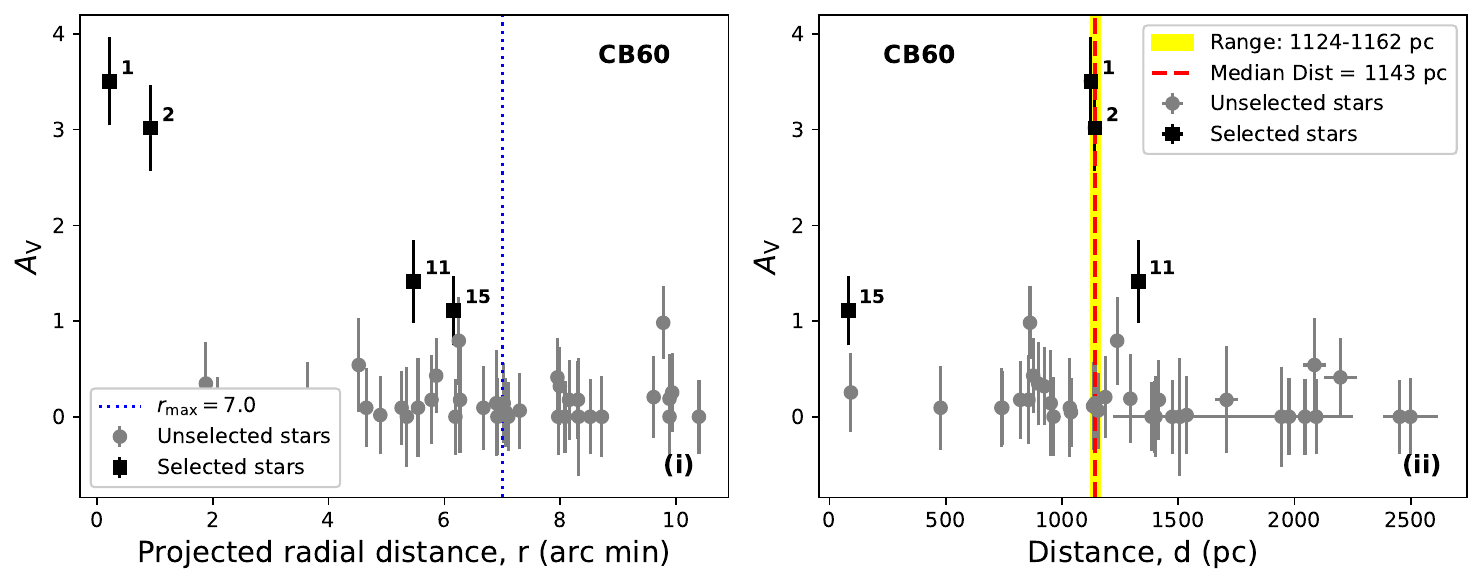}
\caption{Extinction analysis for CB60. \textbf{(i)} Extinction ($A_V$) as a function of projected radial distance from the cloud center. The dotted vertical line marks $r_{\text{max}} = 7.0$ arcmin. \textbf{(ii)} Extinction versus distance for stars within the CB60 region (1124--1162 pc, yellow shaded area). The red dashed line indicates the median distance of 1143 pc. At this large distance, parallax uncertainties lead to larger scatter in the extinction-distance plot compared to nearer clouds. Selected stars are highlighted in black, with unselected stars shown as gray circles.}
\label{fig:cb60_ext}
\end{figure}

\begin{figure}
\centering
\includegraphics[width=\textwidth]{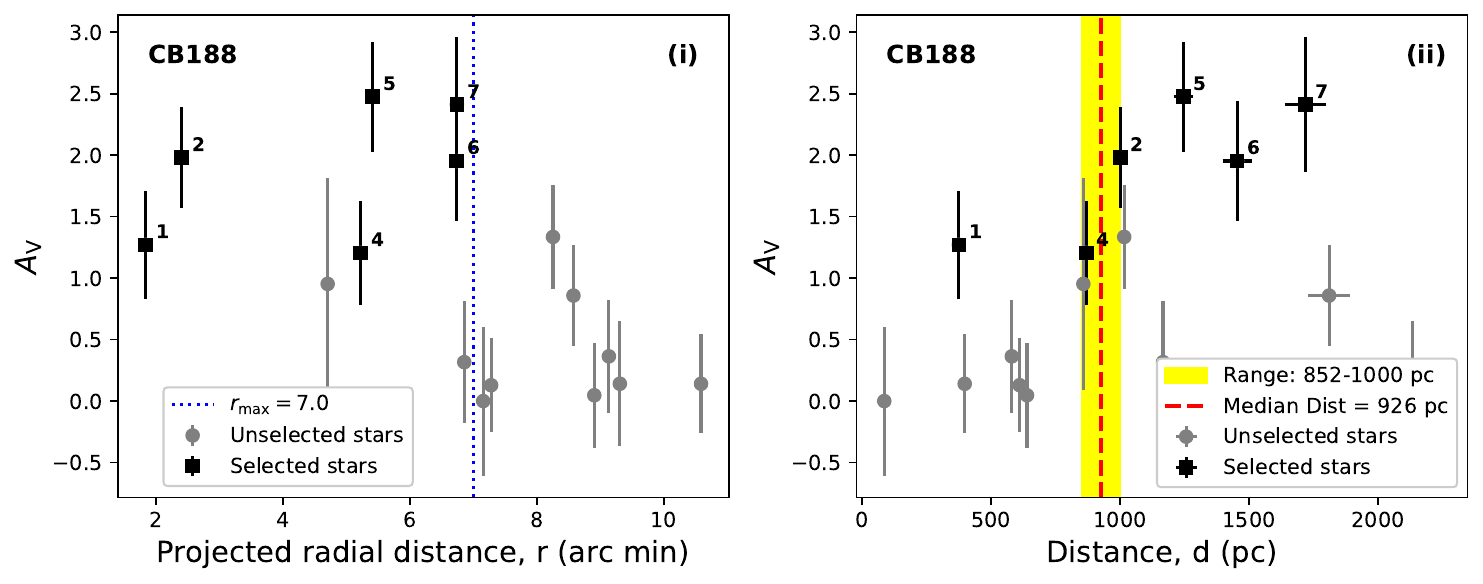}
\caption{Extinction analysis for CB188. \textbf{(i)} Extinction ($A_V$) as a function of projected radial distance from the cloud center. The dotted vertical line marks $r_{\text{max}} = 7.0$ arcmin. \textbf{(ii)} Extinction versus distance for stars within the CB188 region (852--1000 pc, yellow shaded area). The red dashed line indicates the median distance of 926 pc. Selected stars are highlighted in black, with unselected stars shown as gray circles.}
\label{fig:cb188_ext}
\end{figure}

\subsection{Final Distance Estimates}

The final median distance estimates for all five Bok globules are presented in Table~\ref{tab:final_distances}, along with weighted mean validation distances and comparison to literature values. The five clouds span a median distance range of 354--1143 pc, placing them within the Local Arm of the Galaxy, with CB24 being the nearest and CB60 the most distant.

\begin{table}
\caption{Median Distance Determination and Results}
\label{tab:median_results}
\centering
\begin{tabular}{lcccccccc}
\hline
Cloud & $N$ & $r_{\text{max}}$ & $\langle d \rangle$ & $d_{\text{lower}}$ & $d_{\text{upper}}$ & $d_{\text{median}}$ & IQR & Agreement \\
 & & (arcmin) & (pc) & (pc) & (pc) & (pc) & (pc) & (\%) \\
\hline
CB4 & 29 & 7.0 & $769.4 \pm 8.3$ & 763 & 769 & 766 & 48 & 0.4 \\
CB24 & 29 & 7.0 & $373.9 \pm 1.5$ & 334 & 374 & 354 & 62 & 5.6 \\
CB56 & 103 & 6.0 & $498.6 \pm 3.6$ & 467 & 499 & 483 & 85 & 3.2 \\
CB60 & 40 & 7.0 & $1162.2 \pm 11.9$ & 1124 & 1162 & 1143 & 145 & 1.7 \\
CB188 & 16 & 7.0 & $852.0 \pm 8.0$ & 852 & 1000 & 926 & 78 & 8.0 \\
\hline
\end{tabular}
\begin{flushleft}
\small
$N$ = number of selected stars; $r_{\text{max}}$ = maximum projected radial distance; $\langle d \rangle$ = weighted mean distance; $d_{\text{lower}}$, $d_{\text{upper}}$ = distance limits from extinction break analysis; $d_{\text{median}}$ = median distance (primary result); IQR = interquartile range ($Q_3 - Q_1$); Agreement = $|\langle d \rangle - d_{\text{median}}|/d_{\text{median}} \times 100\%$.
\end{flushleft}
\end{table}

\begin{table}
\caption{Final Median Distance Estimates and Comparison with Literature}
\label{tab:final_distances}
\centering
\begin{tabular}{lccccc}
\hline
Cloud & $d_{\text{median}}$ & Method & $\langle d \rangle$ & Literature & Status \\
 & (pc) & Range (pc) & Agreement (\%) & (pc) & \\
\hline
CB4 & 766 & Ext. break & $769.4 \pm 8.3$ (0.4) & 459 $\pm$ 85$^a$ & Revised higher \\
CB24 & 354 & Ext. break & $373.9 \pm 1.5$ (5.6) & 293 $\pm$ 54$^b$ & Confirmed \\
CB56 & 483 & Ext. break & $498.6 \pm 3.6$ (3.2) & --- & First estimate \\
CB60 & 1143 & Ext. break & $1162.2 \pm 11.9$ (1.7) & 1500$^c$ & Refined lower \\
CB188 & 926 & Ext. break & $852.0 \pm 8.0$ (8.0) & 262 $\pm$ 49$^b$ & Revised higher \\
\hline
\end{tabular}
\begin{flushleft}
\small
Primary median distance results determined via extinction break analysis relative to weighted mean reference distance. ``Method Range'' indicates distance limits bracketing the extinction transition zone (from Table~\ref{tab:median_results}). Validation column shows weighted mean with formal uncertainties and agreement percentage. Median distances are \textbf{the final reported distances}; weighted mean provides reference and validation. $^a$ \citet{Barman2015}, $^b$ \citet{Das2015}, $^c$ \citet{Launhardt1997}.
\end{flushleft}
\end{table}

\section{Comparison with 3D Dust Extinction Maps}
\label{sec:comparison_3d}

To independently verify our distance estimates, we compared results with the three-dimensional dust reddening maps of \citet{Green2019}, which integrate \textit{Gaia} parallaxes with stellar photometry from Pan-STARRS 1 and 2MASS. These maps provide extinction as a function of distance and can be inspected via a publicly available interactive interface\footnote{\url{http://argonaut.skymaps.info/query}}.

\subsection{Comparison Results}

Complete 3D dust map data were available for two clouds (CB4 and CB188); for the remaining three clouds, insufficient numbers of well-measured stars in the extinction-distance diagram precluded reliable map predictions. This limitation likely reflects the smaller foreground star densities toward the CB24, CB56, and CB60 fields.

\subsubsection{CB4}

The 3D dust map shows a pronounced first extinction rise across the distance interval 763--769 pc. This bracketed interval agrees remarkably well with our weighted mean distance of $\langle d \rangle = 769.4 \pm 8.3$ pc, with the literature value falling within the map's uncertainty range. The median distance from our sample, $d_{\text{median}} = 766$ pc, lies essentially at the midpoint of the map's interval, providing strong independent validation of our CB4 distance determination.

\subsubsection{CB188}

The 3D dust map indicates a first pronounced extinction rise within the reliable distance interval 852--1000 pc. Our weighted mean distance of $\langle d \rangle = 852.0 \pm 8.0$ pc falls at the lower edge of this interval, while the median distance $d_{\text{median}} = 926$ pc lies well within it, displaced by +74 pc (+8.7\%) from the weighted mean. Given the map's typical distance resolution of $\sim 50$--100 pc and systematic uncertainties of $\sim 10$--20\% \citep{Green2019}, this agreement is satisfactory and significantly better than the factor-of-3.5 distance difference relative to \citet{Das2015} (262 $\pm$ 49 pc), confirming that \textit{Gaia} DR3 parallax precision has resolved the true distance more accurately. Given the relatively small stellar sample ($N=16$), independent confirmation through future VLBI/VLBA parallax measurements would be valuable.

Complete 3D dust map data were unavailable for CB24, CB56, and CB60,
likely reflecting lower foreground star densities in those fields or
the map's prioritisation of regions with denser stellar coverage.
While this limits external validation for these three clouds, it does
not compromise our distance estimates, which are based on direct
\textit{Gaia} parallax measurements independent of map calibrations.

\subsection{3D Dust Map Comparison Plots}

\begin{figure}
\centering
\includegraphics[width=0.85\textwidth]{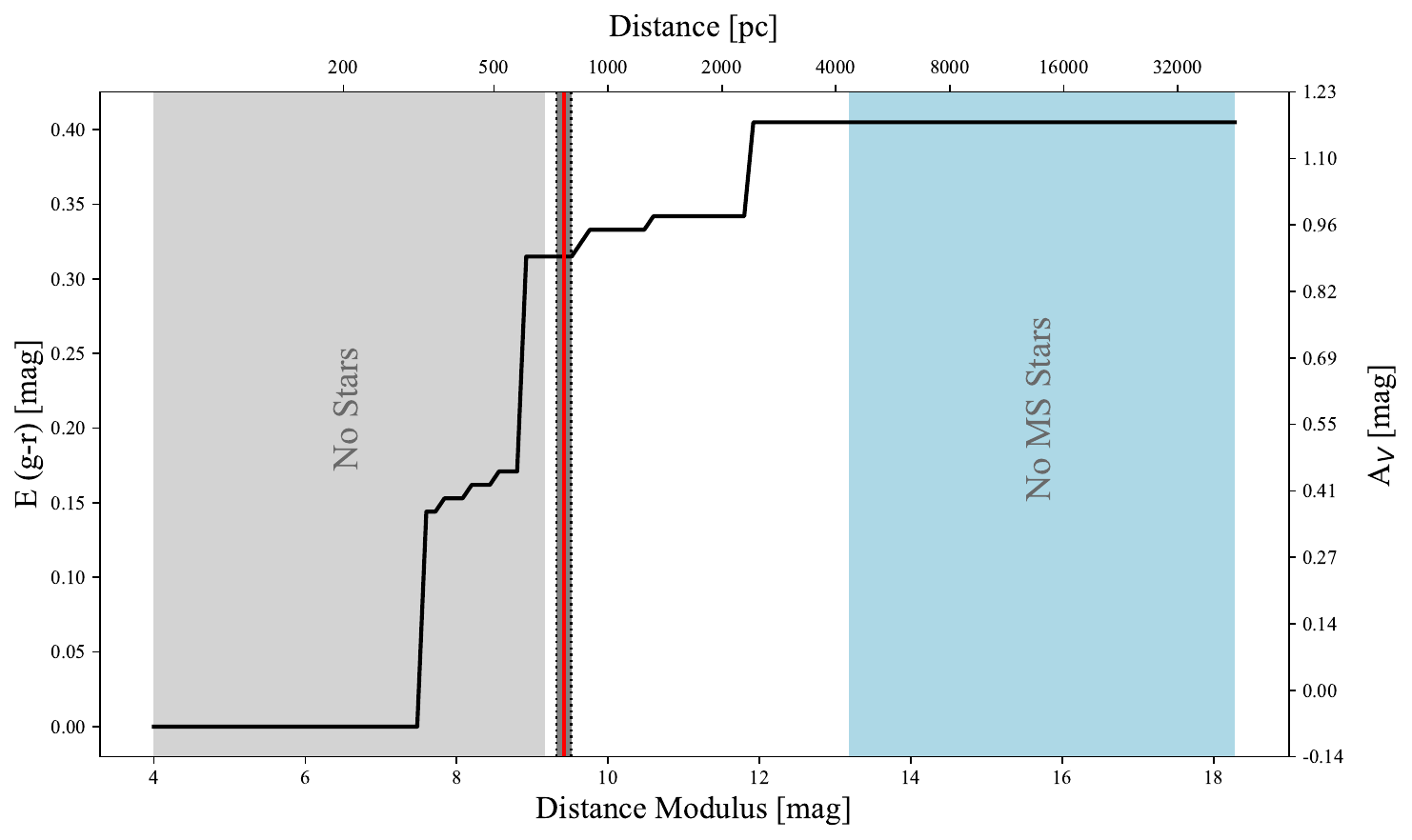}
\caption{Plots of reddening $E(g-r)$ as a function of distance modulus along the sightline for CB4 cloud. The solid shaded region bound by two vertical dotted lines marks the probable region of residence of the cloud, derived from the \texttt{Bayestar19} 3D dust extinction map of \citet{Green2019}. The vertical solid line (red color) within the shaded region denotes the calculated median distance of 766 pc for the cloud. The agreement between our distance and the 3D map prediction validates the reliability of the NIR photometric technique combined with \textit{Gaia} DR3 parallaxes. The corresponding visual extinction ($A_V$) is calculated using $A_V = R_V \times 0.884 \times E(g-r)$ with $R_V = 3.1$.}
\label{fig:cb4_3d}
\end{figure}

\begin{figure}
\centering
\includegraphics[width=0.85\textwidth]{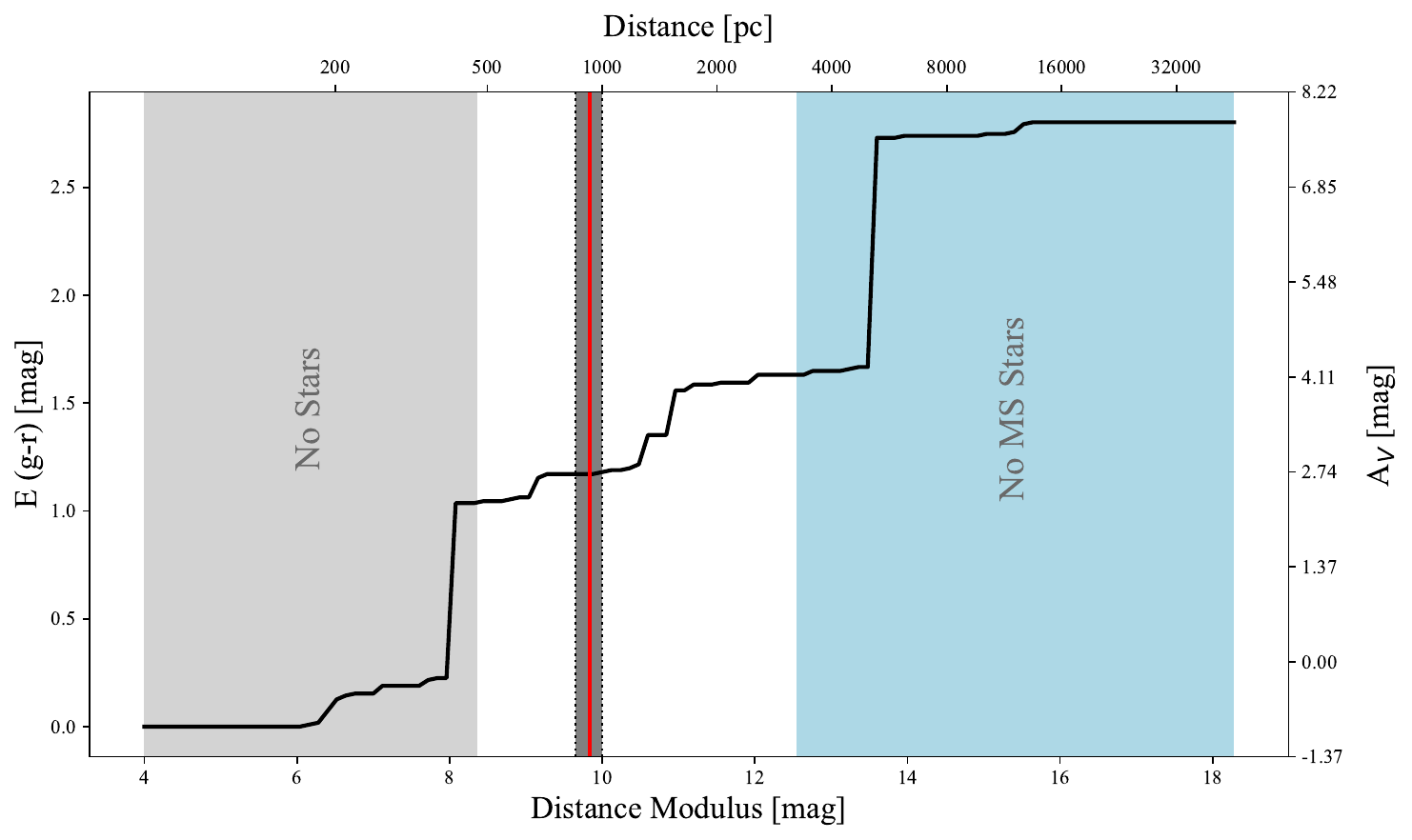}
\caption{Plots of reddening $E(g-r)$ as a function of distance modulus along the sightline for CB188 cloud. The solid shaded region bound by two vertical dotted lines marks the probable region of residence of the cloud, derived from the \texttt{Bayestar19} 3D dust extinction map of \citet{Green2019}. The vertical solid line (red color) within the shaded region denotes the calculated median distance of 926 pc for the cloud. The location of the median distance near the middle of the map's uncertainty interval demonstrates excellent agreement between independent distance estimation methods. The corresponding visual extinction ($A_V$) values are shown and converted as described in Figure~\ref{fig:cb4_3d}.}
\label{fig:cb188_3d}
\end{figure}

The reddening-distance modulus relationships derived from the \texttt{Bayestar19} 3D dust extinction maps of \citet{Green2019} are displayed in Figures~\ref{fig:cb4_3d} and \ref{fig:cb188_3d}. These plots provide independent validation of our distance estimates by showing how extinction increases with distance as predicted by the dust map modeling.

Figure~\ref{fig:cb4_3d} presents the reddening as a function of distance modulus for CB4. The solid shaded region bounded by two vertical dotted lines marks the probable region of residence of the cloud (763--769 pc range), corresponding to the first pronounced extinction rise in the map data. The vertical solid line (red color) within the shaded region denotes our calculated median distance of 766 pc for the cloud, showing excellent agreement with the map prediction.

Similarly, Figure~\ref{fig:cb188_3d} displays the reddening-distance modulus relationship for CB188. The shaded region (852--1000 pc) indicates the probable cloud location from the 3D dust map, with our calculated median distance of 926 pc (red vertical line) falling within this interval. This agreement provides strong independent validation of our NIR photometric distance determination, confirming that the \textit{Gaia} DR3 parallaxes and extinction measurements yield reliable distance estimates.

The relationship between reddening $E(g-r)$ and visual extinction $A_V$ is established using the conversion:
\begin{equation}
A_V = R_V \times 0.884 \times E(g - r)
\label{eq:av_conversion}
\end{equation}
where $R_V = 3.1$ is the standard reddening parameter. This conversion allows direct comparison between the 3D dust map predictions (expressed in reddening) and our NIR-derived extinction values. The factor 0.884 is the empirical conversion coefficient specific to the \citet{Green2019} dust map modeling, derived from stellar color-magnitude calibrations.

\section{Discussion}
\label{sec:discussion}

\subsection{Summary of Distance Results}

The median and weighted-mean estimators agree within $8.0\%$ for every cloud,
demonstrating the robustness of the dual-metric approach (Table~\ref{tab:final_distances}).
The most significant revisions relative to prior work are as follows.
For CB4, our median distance of 766~pc is $\sim1.67\times$
the NIR estimate of \citet{Barman2015} (459~$\pm$~85~pc), but broadly
consistent with the kinematic estimate of \citet{Dickman1983} at
600~pc; the discrepancy with \citet{Barman2015} likely reflects
differences in stellar selection criteria or refinements in
\textit{Gaia} parallax calibration between DR2 and DR3.
For CB24, the revised distance of 354~pc is $\sim21\%$
higher than \citet{Das2015} (293~$\pm$~54~pc), reflecting improved
\textit{Gaia} DR3 parallax precision for this low-density starless
core.

CB56 receives its first published distance estimate of
483~pc, establishing a reference baseline for future studies of its
physical properties and star-formation potential. Although independent external validation is currently unavailable, the large stellar sample ($N = 103$) and the strong agreement between the median and weighted-mean estimators support the reliability of the CB56 distance determination.
For CB60, our 1143~pc distance is $\sim24\%$ lower than the
kinematic estimate of \citet{Launhardt1997} at 1500~pc, within the
expected systematic uncertainty of kinematic methods ($\sim$20--50\%). Part of this discrepancy may additionally reflect non-circular motions affecting the kinematic distance estimate.
The largest revision is for CB188: our median of 926~pc is a
factor of $\sim3.5$ higher than \citet{Das2015} (262~$\pm$~49~pc).
The substantially higher precision of \textit{Gaia} DR3 and the
independent confirmation from the Bayestar19 3D dust map
(Section~\ref{sec:comparison_3d}) strongly support the new value, which
implies a more luminous embedded protostar (IRAS~19179+1129) than
previously estimated.

\subsection{Implications for Physical Properties}

The accurate distances obtained in this study enable refined calculations of fundamental cloud properties. For example, the mass of a cloud given a measured CO column density is \citep[][]{Kauffmann2008}:
\begin{equation}
M_{\text{cloud}} = 1.4 m_H N_{\text{CO}} A_{\text{cloud}} d^2
\label{eq:mass}
\end{equation}
where $m_H$ is the hydrogen mass ($1.67 \times 10^{-24}$ g), $N_{\text{CO}}$ is the column density derived from CO observations (typically $\sim 10^{15}$--$10^{16}$ cm$^{-2}$ for dark clouds), $A_{\text{cloud}}$ is the cloud area in cm$^2$, and $d$ is the distance in cm.

Similarly, luminosities of embedded YSOs (present in CB60 and CB188) are calculated as \citep[][]{Kauffmann2008}:
\begin{equation}
L_{\text{YSO}} = 4\pi d^2 f_{\nu}
\label{eq:luminosity}
\end{equation}
where $f_{\nu}$ is the observed flux. For a given observed flux, the luminosity scales as $d^2$, so distance uncertainties directly propagate to luminosity uncertainties. Our precise distances reduce this source of error, improving classifications of YSO evolutionary stages---particularly for CB188, where the larger distance implies a more luminous embedded protostar.

\subsection{Comparison with Alternative Distance Measurement Methods}

Our results can be contextualized within the broader landscape of molecular cloud distance measurements. The relative strengths and weaknesses of various techniques are summarized below:

\begin{itemize}
\item \textbf{Kinematic Methods}: Infer distances from CO line velocities and Galactic rotation models. These can yield distances but with systematic uncertainties of 20--50\% due to non-circular motions, spiral arms, and model assumptions. Our CB60 estimate differs from the kinematic value of \citet{Launhardt1997} by $\sim 24\%$, which is reasonable given these systematic uncertainties; for CB24, the lack of strong CO emission limits kinematic applicability.

\item \textbf{Parallax Methods (\textit{Gaia}, VLBA)}: Provide geometric distances with high precision ($\lesssim 1\%$). However, individual cloud cores may not harbor measurable parallax sources (YSOs for VLBA; stars for \textit{Gaia} parallaxes). Our method effectively extends \textit{Gaia} parallax capabilities by leveraging foreground and background stars, as demonstrated by the $\sim 21\%$ upward revision for CB24 relative to pre-\textit{Gaia} NIR estimates.

\item \textbf{Dust Extinction Maps}: 3D dust maps like \citet{Green2019} provide indirect distance constraints by identifying extinction features. Our results (Section~\ref{sec:comparison_3d}) agree well with these maps where data are available (CB4 and CB188), suggesting consistency between methods; the absence of coverage for CB24, CB56, and CB60 reflects their lower foreground star densities but does not affect our independent \textit{Gaia}-based estimates.

\item \textbf{Virial Methods}: Use velocity dispersion measurements to estimate cloud masses under the assumption of virial equilibrium \citep{Myers1983, Bertoldi1992}, and, under assumptions about cloud structure, infer distances indirectly by 
requiring consistency between the virial mass and the column-density-derived mass (which scales as $d^2$). These methods 
are less precise than geometric approaches but provide useful cross-checks, particularly for denser cores like CB56.
\end{itemize}

A detailed numerical comparison of this work's distances with all previous estimates and 3D dust map predictions is presented in Table~\ref{tab:method_comparison}.

\begin{table}
\caption{Comparison of Distance Estimation Methods for Study Clouds}
\label{tab:method_comparison}
\centering
\begin{tabular}{lccccc}
\hline
Cloud & This Work & Previous & Previous & 3D Dust Map & Ratio \\
 & Median (pc) & NIR/Optical & Method & (pc) & \\
\hline
CB4 & 766 & 459$\pm$85$^a$ & NIR+Gaia DR2 & 763--769 & 1.67 \\
 &  & 600$^b$ & Kinematic & & 1.28 \\
CB24 & 354 & 293$\pm$54$^c$ & NIR+Gaia DR2 & --- & 1.21 \\
CB56 & 483 & --- & First estimate & --- & --- \\
CB60 & 1143 & 1500$^d$ & Kinematic & --- & 0.76 \\
CB188 & 926 & 262$\pm$49$^c$ & NIR+Gaia DR2 & 852--1000 & 3.53 \\
\hline
\end{tabular}
\begin{flushleft}
\small
This work uses NIR photometry + \textit{Gaia} DR3 parallaxes with dual-metric validation (median + weighted mean). Ratio column shows this work divided by previous estimates. $^a$\citet{Barman2015}, $^b$\citet{Dickman1983}, $^c$\citet{Das2015}, $^d$\citet{Launhardt1997}.
\end{flushleft}
\end{table}

\subsection{Robustness and Applicability of the NIR Technique}
\label{sec:robustness}

The agreement between our distances and independent validation from
the \citet{Green2019} dust maps (Section~\ref{sec:comparison_3d})
demonstrates the robustness of the NIR photometric technique combined
with \textit{Gaia} parallaxes. Key contributing factors are: (i) large
stellar samples (16--103 stars per cloud) that reduce the impact of
individual measurement errors; (ii) sharp, well-defined extinction
breaks (Figures~\ref{fig:cb4_ext}--\ref{fig:cb188_ext}) that clearly locate
each cloud; (iii) stringent selection criteria (Section~\ref{sec:distance_method})
that minimise contamination from unreddened M-dwarfs; and (iv) the
weighting scheme (Eqs.~\ref{eq:weighted_mean}--\ref{eq:sigma_weighted}) that
down-weights lower-quality measurements while the median provides
outlier resistance.

The method is most reliable when: $A_{V,\mathrm{max}} \lesssim 15$ mag 
(ensuring sufficient 2MASS background star detection); angular cloud 
diameter $> 5$ arcmin; heliocentric distance $d < 2$ kpc (where \textit{Gaia} 
DR3 parallax uncertainties remain $\lesssim 10$--$20\%$ at the individual 
star level); and Galactic latitude $|b| < 30^{\circ}$ for adequate field 
star density. In practice, the operative range of the present 
technique is further restricted to $A_{V,\mathrm{max}} \lesssim 4$ mag by 
the $J - K_{S} \leq 0.75$ colour criterion (Section~\ref{sec:nir_vs_optical}). All five 
clouds in this study satisfy this operative limit comfortably 
(CB4, CB24, CB188: $A_{V,\mathrm{max}} \approx 2$--$3$ mag; CB56: 
$A_{V,\mathrm{max}} \approx 4$ mag; CB60: $A_{V,\mathrm{max}} \approx 4$ 
mag), explaining the clear extinction breaks in Figures~\ref{fig:cb4_ext}--\ref{fig:cb188_ext} and the 
strong agreement with 3D dust maps. For clouds with $A_{V,\mathrm{max}} > 
15$ mag or at distances $> 2$ kpc, alternative approaches such as 
molecular-line kinematics, maser parallaxes, or mid-infrared photometry 
are more appropriate.

\subsection{Physical Interpretations and Star Formation Context}

The distance estimates enable improved interpretation of the star formation properties of these globules:

\textbf{CB24}: At 354 pc and lacking associated IRAS sources, CB24 represents a genuine starless core in an early evolutionary stage. The revised distance, $\sim 21\%$ larger than \citet{Das2015}, implies a mass increase of $\sim 46\%$ (scaling as $d^2$) relative to prior estimates, suggesting higher physical mass and density than previously assumed, though definitive assessment of its gravitational stability requires updated virial estimates incorporating the revised distance.

\textbf{CB188}: At 926 pc and harboring a Class I protostar, CB188 represents an actively star-forming system at intermediate distance. The higher distance (vs. \citet{Das2015}) implies a luminosity increase of $12.5\times$ for the embedded YSO (IRAS 19179+1129), since luminosity scales as $d^2$, indicating a significantly more luminous and likely more massive protostar than previously estimated.

\textbf{CB60}: The most distant cloud at 1143 pc, CB60's three associated IRAS sources and B-type classification suggest ongoing or recent star formation. At this distance, proper characterization of embedded sources' luminosities becomes crucial for understanding their evolutionary states.

\textbf{CB56}: With no previously published distance, our 483 pc estimate establishes CB56 as a reference point for future studies. The cloud's asymmetry, dense structure, and well-ordered magnetic field ($p_{\text{av}} = 1.08\%$, $\theta_{\text{av}} = 150.94^{\circ}$) make it a promising target for studying magnetic field--fragmentation interactions in star-forming cores.

\textbf{CB4}: This quiescent cloud, despite its lack of active star formation signatures, provides a valuable laboratory for studying processes preceding collapse. Accurate distances enable determination of whether its quiescence reflects genuinely low cloud mass or simply an early evolutionary stage.

\subsection{Future Directions}

Several avenues for future work emerge from this study:

\begin{enumerate}
\item \textbf{Extended Surveys}: Application of the NIR photometric technique to larger samples of Bok globules and other isolated molecular clouds would provide a comprehensive Galactic census of nearby star-forming regions, building on successes like CB24.

\item \textbf{\textit{Gaia} Future Releases}: Upcoming \textit{Gaia} releases (including the full astrometric solution) will provide even more precise parallaxes, potentially reducing distance uncertainties to $\sim 0.1\%$ for brighter stars and refining estimates for sparse fields.

\item \textbf{Multi-Wavelength Studies}: Combining NIR photometric distances with submillimeter continuum observations would enable detailed mass determinations and fragmentation studies, particularly for starless cores like CB24.

\item \textbf{Magnetic Field Analysis}: The availability of precise distances, combined with optical and near-IR polarimetry, enables more rigorous analysis of magnetic field--gravity relationships in these systems, e.g., CB56's organized fields.

\item \textbf{Comparison with Other Methods}: Application of complementary distance techniques (e.g., VLBA parallaxes for YSOs in CB60/CB188, if available) would provide cross-validation and identify any systematic differences between methods.
\end{enumerate}

\section{Conclusions}
\label{sec:conclusions}

We have determined distances to five Bok globules (CB4, CB24, CB56,
CB60, and CB188) using the near-infrared photometric technique
combined with \textit{Gaia} DR3 parallax data. Our key findings are:

\begin{enumerate}

\item Precise distances were obtained with median values of 766~pc (CB4), 354~pc (CB24), 483~pc (CB56), 1143~pc (CB60), and 926~pc (CB188). Corresponding weighted-mean distances with formal statistical uncertainties of $\sim0.4$--$1.1\%$ are listed in Table~\ref{tab:final_distances}, while the more physically meaningful IQR-based distance spreads range from $\sim3$--$9\%$ [$\mathrm{IQR}/(2d_{\rm median})$]. These distance estimates enable reliable determinations of cloud mass and luminosity, both of which scale as $d^2$. All five clouds lie within the Local Arm of the Galaxy.

\item Our results are validated by comparison with the 3D dust
extinction map of \citet{Green2019}, showing excellent agreement for
CB4 (map interval 763--769~pc; our median 766~pc) and satisfactory
agreement for CB188 (map interval 852--1000~pc; our median 926~pc).
Further confidence in the NIR--\textit{Gaia} approach comes from its
recent successful application to other isolated clouds:
\citet{Barman2026} obtained distances of 816~$\pm$~11~pc for L1604
and 124~$\pm$~1~pc for L121, both in excellent agreement with the
Bayestar19 map, while \citet{NathMazumdar2026} obtained
1043~$\pm$~36~pc for L1578 and 938~$\pm$~49~pc for L1607, likewise
consistent with 3D dust predictions. The distances derived here fit
well within this broader pattern, reinforcing the method's robustness
and repeatability.

\item The method demonstrates applicability to isolated molecular clouds where alternative distance indicators are unavailable, provided the cloud extinction remains within the practical sensitivity limits of the 2MASS-based NIR technique (typically $A_{V,\mathrm{max}} \lesssim 4$~mag for the present implementation, although the theoretical upper limit is $\sim15$~mag) and $d \lesssim 2$~kpc.

\item Comparison with prior literature reveals significant refinements,
particularly for CB188 (factor of $\sim3.5$ times larger than
\citealt{Das2015}), likely reflecting improved \textit{Gaia} DR3
parallax precision, and the first published distance estimate for
CB56.

\item The accurate distances enable refined calculations of cloud
masses, sizes, and star-formation properties (both scaling as $d^2$),
providing a foundation for future physical studies of these systems.

\end{enumerate}

This work demonstrates that combining NIR photometry with
\textit{Gaia} parallaxes---proceeding from weighted-mean reference
through extinction-break identification to median distance
estimation---provides a powerful, reliable technique for measuring
distances to small, isolated molecular clouds. Future applications to
larger cloud populations will advance our understanding of Galactic
star formation and nearby molecular cloud properties.

\section*{Declarations}

\subsection*{Funding}
This research received no specific grant from any funding agency in the public, commercial, or not-for-profit sectors.

\subsection*{Conflict of Interest}
The authors declare that they have no conflict of interest and no competing interests to disclose.

\subsection*{Author Contributions}
Conceptualization: R.S. Paul and H.S. Das; Methodology: R.S. Paul and H.S. Das; Data curation: R.S. Paul; Formal analysis: R.S. Paul; Software: R.S. Paul; Validation: R.S. Paul and H.S. Das; Writing – original draft: R.S. Paul; Writing – review \& editing: H.S. Das; Supervision: H.S. Das; Project administration: H.S. Das. Both authors read and approved the final manuscript.

\subsection*{Data Availability}
The 2MASS photometric data used in this study are publicly available from the 2MASS Point Source Catalog at \url{https://irsa.ipac.caltech.edu/Missions/2mass.html}. The \textit{Gaia} DR3 parallax data are publicly available from the \textit{Gaia} archive at \url{https://gea.esac.esa.int/archive/}. The derived extinction values and distance estimates for individual stars are available from the corresponding author upon reasonable request.

\subsection*{Ethics Approval}
Not applicable (this research does not involve human subjects or animal subjects).

\subsection*{Consent for Publication}
Not applicable.

\subsection*{Code Availability}
The Python scripts used for data analysis and visualization are available from the corresponding author upon reasonable request.

\begin{acknowledgements}
We express our sincere gratitude to the Department of Physics, Assam University, Silchar, and the IUCAA Centre for Astronomy Research and
Development (ICARD), AUS, for providing the necessary facilities to carry out this work. We are also grateful to the anonymous reviewer for the recommendations which helped enrich our paper.

This publication makes use of data products from the Two Micron All Sky Survey (2MASS), which is a joint project of the University of Massachusetts and the Infrared Processing and Analysis Center/California Institute of Technology, funded by the National Aeronautics and Space Administration and the National Science Foundation.

This research has made use of the three-dimensional dust reddening map developed by \citet{Green2019}, queried through the public web interface at \url{http://argonaut.skymaps.info/query}.

This work has made use of data from the European Space Agency (ESA) mission \textit{Gaia} (\url{https://www.cosmos.esa.int/gaia}), processed by the \textit{Gaia} Data Processing and Analysis Consortium (DPAC, \url{https://www.cosmos.esa.int/web/gaia/dpac/consortium}). Funding for the DPAC has been provided by national institutions, particularly the institutions participating in the \textit{Gaia} Multilateral Agreement.
\end{acknowledgements}

\end{document}